\documentclass[prl,aps,amsmath,amssymb,reprint]{revtex4-1}
\usepackage{graphicx}
\usepackage{dcolumn}
\usepackage{bm}
\newcommand{\p}{$\%$}
\newcommand{\pat}{{${~at.}$}\%}
\newcommand{\pn}{$\mathrm{R{_{N_2}}}$}

\newcommand{\tcn}{$\mathrm{Co_{4}N}$}

\newcommand{\muB}{$\mathrm{\mu_{B}}$}

\newcommand{\tfn}{$\mathrm{Fe_{4}N}$}
\newcommand{\tnn}{$\mathrm{Ni_{4}N}$}
\newcommand{\Ts}{$\mathrm{T_{s}}$}

\newcommand{\hf}{$\Delta$$H^{\circ}_f$}

\begin{document}
	\title{Study of Reactively Sputtered Nickel Nitride Thin Films}
	\author{Nidhi Pandey$^1$} \author{Mukul Gupta$^{1}$}\email{mgupta@csr.res.in} \author{Jochen Stahn$^{2}$}
	\address{$^1$UGC-DAE Consortium for Scientific Research, University Campus, Khandwa
	Road, Indore 452 001, India\\$^2$Laboratory for Neutron Scattering and Imaging, Paul Scherrer Institut, CH-5232 Villigen PSI, Switzerland}

	\date{\today}
	
	
	\begin{abstract}
	Nickel nitride (Ni-N) thin film samples were deposited using
	reactive magnetron sputtering process utilizing different partial
	flow of N$_2$ (\pn). They were characterized using x-ray
	reflectivity (XRR), x-ray diffraction (XRD) and x-ray absorption
	near edge spectroscopy (XANES) taken at N $K$-edge and Ni
	$L$-edges. From XRR measurements, we find that the deposition rate
	and the density of Ni-N films decrease due to successively
	progression in \pn, signifying that Ni-N alloys and compounds are
	forming both at Ni target surface and also within the thin film
	samples. The crystal structure obtained from XRD measurements
	suggest an evolution of different Ni-N compounds given by: Ni,
	Ni(N), Ni$_4$N, Ni$_3$N, and Ni$_2$N with a gradual rise in \pn.
	XANES measurements further confirm these phases, in agreement with
	XRD results. Polarized neutron reflectivity measurements were
	performed to probe the magnetization, and it was found Ni-N thin
	films become non-magnetic even when N incorporation increases
	beyond few \pat. Overall growth behavior of Ni-N samples has been
	compared with that of rather well-known Fe-N and Co-N systems,
	yielding similarities and differences among them.     
	\end{abstract}
	
	\maketitle

\section{Introduction}
\label{1}

The family of transition metal nitride (TMN) exhibits interesting
electronic, optical, thermal and magnetic properties. In
particular, the combination of various properties of 3$d$ TMNs
such as wear and corrosion resistant, exceptional hardness with
excellent magnetic properties have attracted considerable
attention~\cite{TMN_NatMat19_Sun_Map,Khazaei_2013_TMN,TMN_Zhong16,TMN_Oyama_96,TMN_Williams_97,JMMM90_Mag_Nit_Coey}.
However among those, early 3$d$ TMNs (e.g. ScN, TiN, VN, and CrN)
are well established and relatively more explored than late 3$d$
TMNs (e.g. Mn-N, Fe-N, Co-N, and Ni-N).

For the early 3$d$ TMNs, mostly the MN (M = metal) stoichiometry
is prevalent but some reports of M$_2$N phase has also observed
e.g. Cr$_2$N and
Nb$_2$N~\cite{TMN_Zhong16,TMN_Ningthoujam_Gajbhiye_15,MG_book_Fe-N}.
On the other hand, a significant change in stoichiometry is
observed as M$_3$N and M$_4$N, etc. for late 3$d$ TMNs. The
increased M/N ratio for the late TMNs signifies the rejection of N
by the metal atoms which reflected in their poor stability and
challenging formation~\cite{TMN_Zhong16,TMN_Oyama_96,TMN_Ningthoujam_Gajbhiye_15}. Nevertheless, the late 3$d$ TMNs are
largely investigated for their excellent electronic and magnetic
properties~\cite{JMMM90_Mag_Nit_Coey,PRB18_Theo_Fe4N/Co4N/MgO,FeN_Review_Bhattacharyya15,FeX_N_B_JMMM18_Hui18,Fe4N_Blanca09_Review,JPCM_Markus16,JMMM10_Imai,JAC14_Imai}.

Unlike early, the late 3$d$ TMNs have been known to possess
several multi-nitride phases. For instance, in Fe-N, different
crystallographic phases have been obtained:
Fe$_{16}$N$_2$~\cite{Fe16N2_Metzger94},
Fe$_8$N~\cite{Fe8N_Matar92}, \tfn~\cite{Fe4N_Blanca09_Review},
Fe$_3$N~\cite{Fe3N_Leineweber_JAC99}, Fe$_2$N~\cite{Fe2N_JAC18},
FeN~\cite{FeN_Zhao_RSC15}, FeN$_2$, FeN$_4$,
FeN$_8$~\cite{FeNx_CoNx_Niwa_ACS17,FeNx_Bykov_Nat18,FeNx_Lailei_ACS18}.
Similarly, in Co-N system,
\tcn~\cite{APA19_Co4N_NP,JMMM18_Co4N_NP,JAC16_NPandey,PRB19_Co4N_NP},
Co$_3$N~\cite{Vac01_Asahara},
Co$_2$N~\cite{JVSTA04_Fang,EupChem17_Co2N_Zhou,ASS19_Co2N_Guo_ORR},
CoN~\cite{JAC95_Suzuki,AFM08_CoN} and
CoN$_2$~\cite{FeNx_CoNx_Niwa_ACS17} phases have been synthesized.
On the other hand, the nickel nitrides (Ni-N) is scariest among
the late 3$d$ TMNs as only a handful of reports are available.
However, Ni-N are potentially important metallic
compound~\cite{Ni-N_nanosheet_metallic_18,Ni3N_Electrode_JMCA15}
as they have been reported to serve as negative electrodes in
lithium batteries and energy storage
devices~\cite{Ni-N_GaN_anode_Li_Vac19,Ni-N_lithium_electrodes_Gillot11,
	Ni3N_Electrode_JMCA15,TMN_electrode_Balogun_RSC15,FMS19_NiO_Ni2N_Lit_ion_Bat},
dye or quantum dot solar
cells~\cite{Ni-N_dye_electrode_Kang_SciRep15,Ni-N_dcMS_Keraudy_TSF19}.
Ni-N compounds have been also found to exhibit superior
(electro)catalytic activities in the reduction reactions and also
demonstrated many other advantages over pure
metals~\cite{Ni-N_nanosheet_metallic_18,Ni3N_Catalytic_Gage16}.
Non-conventional insulating-metal transition properties have also
been demonstrated by
Ni-N~\cite{Ni-N_metal_trans_Kim_IEEE12,Ni-N_metal_trans_JAP14,Ni-N_film_metal_trans_JVST13,Ni-N_metal_trans_ASP14}.
With a good quality of interface, Ni-N demonstrates a high work
function, low leakage current and therefore implemented as an
electrode in GaN-based Schottky barrier
diodes~\cite{Ni-N_GaN_anode_Li_Vac19}.

Notwithstanding the active experimental research work, relatively
less information about the intrinsic properties of Ni-N is
available. Even the phase-diagram information of Ni-N is ambiguous
or
scarce~\cite{Ni-N_thermal_stab_phase_dia_Guillermet1991,Phase_diagram_N_imp_Ni_Neklyudov04}.
Few phases in Ni-N system as;
\tnn(fcc)~\cite{Ni4N_2_Terao59,Ni4N_2_Dorman_TSF83,Ni4N_allotropes_Hemzalova_PRB13,TM4N_2_Fang_RSC14,Ni4N_Leineweber19},
Ni$_3$N(hcp)~\cite{JAC09_Ni3N_film_Vempaire,JAP09_Ni3N_film_Vempaire},
Ni$_2$N~\cite{PhyB14_Ni2N_film_Nishihara,
	Ni-N_Houari_Matar18,Mat_Lat18_Ni2N_Zhi_yuan} have been reported to
synthesized by chemical vapor deposition, sputtering and highly
reactive azides or hydrazine. While some other phases Ni$_3$N$_2$
and NiN$_6$, have also been reported to crystallize by chemical
route methods~\cite{Ni-N_thermal_stab_phase_dia_Guillermet1991}.
On the other hand, the stoichiometric NiN has not been achieved
yet. Recently, nickel pernitride (NiN$_2$) phase is also
synthesized at very high pressure of about
40\,GPa~\cite{NiN2_InorgChem19}. Additionally, an ambiguity about
the magnetic characteristic of Ni-N phases can also be seen from
the available literature. Gajbhiye $et~al.$ reported a
ferromagnetic state in the Ni$_3$N phase with Curie temperature of
about 634\,K~\cite{Gajbhiye2002_Ni-N}. On the other hand, the
Ni$_3$N phase (or mixed with Ni$_2$N phase) was found to exhibit
paramagnetic behavior, in some other
reports~\cite{JAC09_Ni3N_film_Vempaire,JAP09_Ni3N_film_Vempaire,ASS09_Popovic_Ni3N}.
Furthermore, the detailed information on the structural and the
magnetic behavior of Ni-N thin films is still lacking. The dearth
of such information about the Ni-N system is due to their
metastability which is related to their high enthalpy of formation
(\hf) and makes it rather difficult to
synthesize~\cite{Ni-N_thermal_stab_phase_dia_Guillermet1991}. This
indicates that a relatively narrow window exists for the synthesis
of various Ni-N phases.

In view of this, we synthesized the entire spectrum of Ni-N films
with the successive increase of the partial flow of nitrogen (in
the window 0 to 100\p) by reactive magnetron sputtering and
investigated their structural and magnetic properties.

\section{Experimental Procedure} \label{expe}
A set of Ni-N thin films were prepared on amorphous quartz
(SiO$_2$) substrates using direct current magnetron sputtering
system (Orion-8, AJA Int. Inc.) at room temperature. High purity
Ni target (99.993\p~pure) $\phi$~1\,inch was sputtered in the
presence of different partial gas flow of nitrogen ({\pn} =
p$_{\mathrm{N}_2}$/(p$_{\mathrm{Ar}}$+p$_{\mathrm{N}_2}$), where
p$_{\mathrm{Ar}}$ and p$_{\mathrm{N}_2}$ are gas flow of Ar and
N$_2$ gases, respectively). \pn~was varied at 0, 5, 10, 15, 20,
50, 75, and 100\p. A base pressure of 1$\times$10$^{-7}$\,Torr~was
achieved before deposition and the working pressure was maintained
at 3\,mTorr~during deposition. No intentional substrate heating
was provided during and after deposition.

The density and deposition rates were measured using x-ray
reflectivity (XRR) measurements using Cu K-$\alpha$ x-ray source.
Samples were characterized for their crystal structure and phase
formation by x-ray diffraction (XRD) using a standard x-ray
diffractometer (Bruker D8 Advance) using Cu K-$\alpha$ x-ray
source. X-ray absorption near edge spectroscopy (XANES)
measurements were performed in the total electron yield (TEY) mode
at BL-01 at the Indus-2 synchrotron radiation source at RRCAT,
Indore~\cite{XAS_beamline}. Polarized neutron reflectivity (PNR)
measurements were performed at AMOR, SINQ, PSI Switzerland in time
of flight mode using Selene optics~\cite{Amor16_cite_1,Amor17_cite_2}. During PNR measurements, to saturate the sample
magnetically, a magnetic field of 0.5\,T was applied parallel to
the sample surface.

\section{Results and Discussion}
\subsection{Phase formation and electronic structure}

\begin{figure} \center
	\vspace{-1mm}
	\includegraphics [scale=0.23] {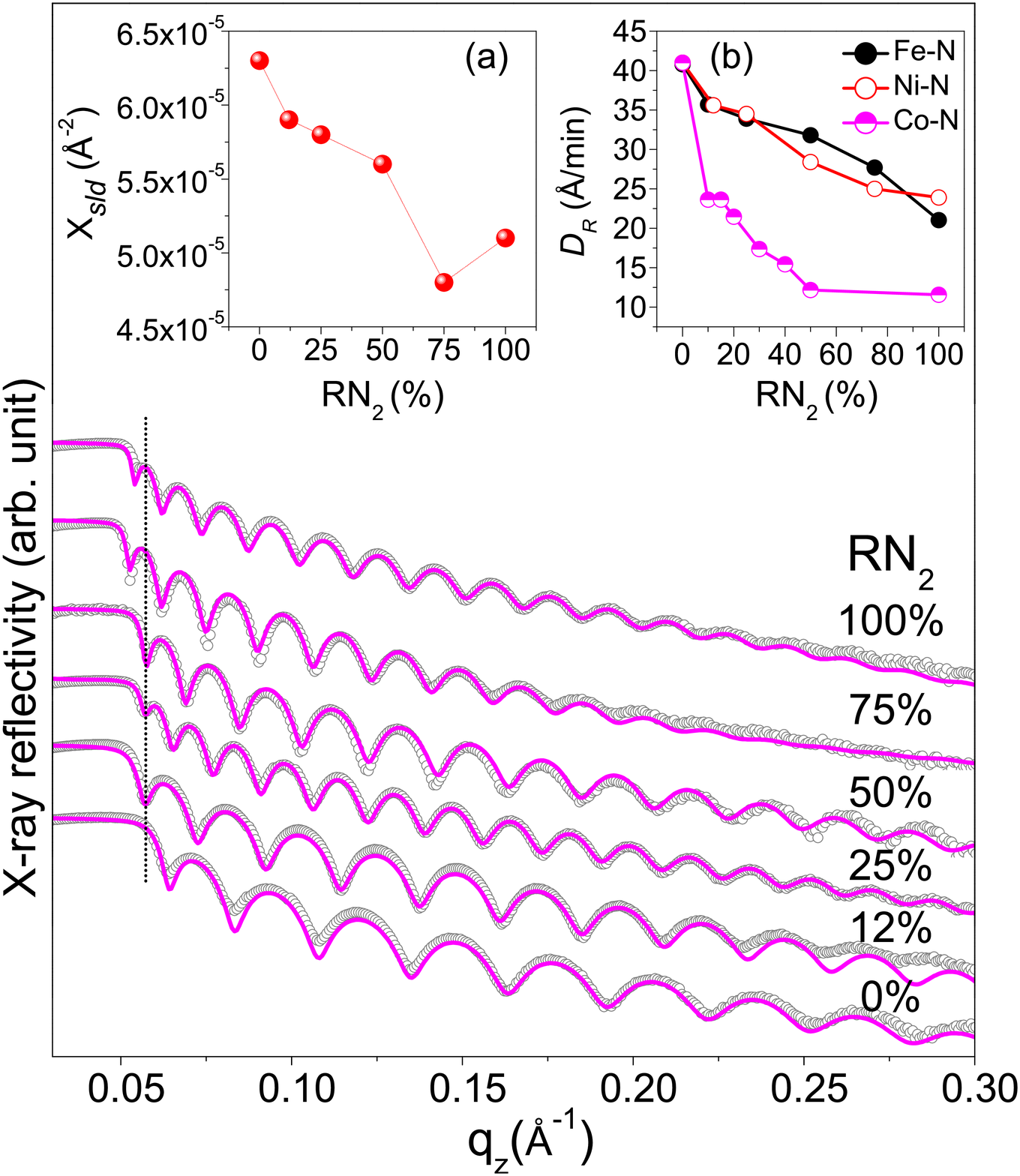}
	\caption{\label{fig:xrr} Fitted XRR patterns of Ni-N thin films
		prepared at successively increasing values of \pn~at \Ts~= 300\,K.
		Inset shows the variation of scattering length density (X$_{sld}$)
		(a) and deposition rate ($D_R$) (b) as a function of \pn. The
		dashed line is a guide to the eye.} \vspace{-1mm}
\end{figure}

XRR patterns of Ni-N thin films deposited at different \pn~are
shown in the fig.~\ref{fig:xrr}. The thickness, roughness and
density of these films have been extracted from the fitting of the
XRR patterns using Parratt32 software~\cite{Parratt32}. We can see
from the XRR patterns that with an increase in \pn, the critical
edges shift towards the lower value (shown by dashed line) which
indicates a gradual drop in scattering length density (X$_{sld}$)
due to the increased incorporation of nitrogen in Ni films as
shown in the inset (a) of fig.~\ref{fig:xrr}. From the
table.~\ref{table}, a significant reduction in X$_{sld}$ (compared
to the pure Ni film) can also be clearly seen. Similar behavior of
X$_{sld}$ with \pn~has also been previously obtained for Co-N
films (not shown here)~\cite{CoN_AIP_Adv2015}. However, roughness
remain nearly constant at about 5\,{\AA}. Apart from this, the
deposition rates ($D_R$) have been measured using the thickness
and the deposition time and shown in the inset (b) of
fig.~\ref{fig:xrr}. For comparison, the normalized $D_R$ obtained
from the Fe-N and Co-N films are also included.

A reduction in $D_R$ with an increase in \pn~can be seen in Ni-N
films similar to that in Fe-N and Co-N obtained using the same
sputtering system with similar $\phi$ 1-\,inch~Fe and Co targets.
Such behavior clearly indicates that some nitride formation is
also taking place at target itself. Such compound formation in a
reactive sputtering process is generally referred as `target
poisoning'. However, it is contrary to a previous report which
ruled out the possibility of target poisoning during the formation
of Ni-N thin films~\cite{Ni-N_fil_Ts_Kawamura_Vac2000}. Here, both
Ni-N and Fe-N systems seem to follow a similar variation in $D_R$
with \pn, whereas the Co-N system shows a rather different
behavior. The $D_R$ reduces about 16 to 42\p~respectively, at
\pn~= 25 to 100\p~compared to the $D_R$ of pure metallic state
(i.e. \pn~= 0\p; $D_R$ = 41\,{\AA}/min) whereas, it drops at about
42 to 72\p~for Co-N. In addition, the poisoned state achieve at
relatively higher \pn~in Ni-N and Fe-N system compared to Co-N. On
comparing the behavior of D$_R$ in Ni-N, Fe-N system with the
Co-N, it is interesting to note that the Co target is more prone
to nitride formation at the target.

\begin{figure} \center
	\vspace{-1mm}
	\includegraphics [scale=0.16] {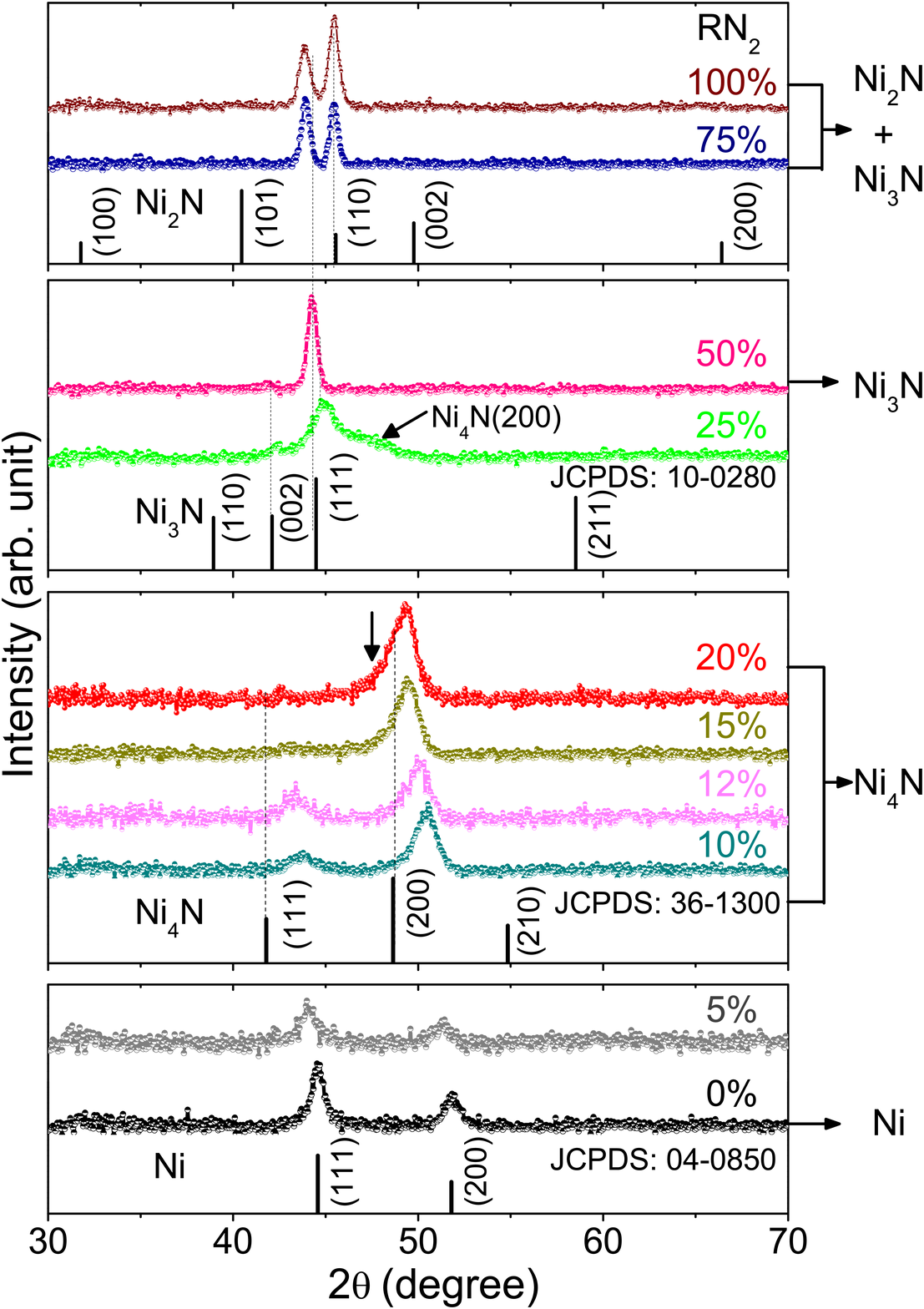}
	\caption{\label{fig:xrd} XRD patterns of Ni-N thin film prepared
		at different values of \pn. For reference, XRD patterns of
		corresponding JCPDS no. are also included. For Ni$_2$N phase,
		theoretically calculated diffraction pattern is given.}
	\vspace{-1mm}
\end{figure}

Using $D_R$ obtained from XRR measurements, a separate set of
samples with a thickness of about 100\,nm was prepared. The XRD
patterns of these samples are shown in fig.~\ref{fig:xrd}. Pure Ni films formed with \pn~=~0\p~exhibit peaks at 44.58 and 51.85$^{\circ}$ correponding to fcc Ni ({JCPDS No. 040850}.
On increasing the \pn~to 5 and 12\p, (111) and (200)
peaks gradually shift to lower 2$\theta$ values and also becomes
slightly border, signifying an interstitial incorporation of N
atom into fcc Ni lattice and can be assigned as Ni(N). Such
behavior is usually achieved in the initial nitride formation of
films as previously also seen in the Fe-N, Co-N and Cr-N systems.
The incorporation of N into Ni lattice is also evident from the
noticeable reduction in the $D_R$ (inset (a) of
fig.~\ref{fig:xrr}) and X$_{sld}$ (table.~\ref{table}). On further
increasing the \pn~to 15\p, the structure changes into \tnn~phase
with preferential orientation of (200) plane. However, the
\tnn~phase formation remains constant up to \pn~= 20\p~but with a
shift towards lower angles in both (111) and (200) peaks,
indicates a further expansion in \tnn~phase with lattice parameter
(LP) about 3.692 $\pm$ 0.005\,{\AA}. However, the obtained LP is
still approximately 1\p~less than the theoretical value
(3.732\,{\AA})~\cite{JPCM_Markus16}. In addition, it may also be
noted here that an asymmetry appears on the onset of (200) peak
(shown by an arrow in fig.~\ref{fig:xrd}) which may be related to
deformation in the cubic structure. On further increasing the
\pn~to 25\p, the \tnn~phase is accompanied by the Ni$_3$N phase.
Whereas it completely get transforms into N rich phase, identified
as hcp Ni$_3$N for \pn~= 50\p. Here, a strong reflection at about
44.25$^{\circ}$ indicates a preferential orientation with the
(111) direction normal to the surface. However, a reactively
sputtered Ni$_3$N film has previously been obtained with preferred
orientation along (002) direction. Such differences in the
preferred orientation directions can be due to high substrate
temperature (475\,K) and higher deposition rate ($\approx$
2.9\,{\AA}/sec) than the present work~\cite{Ni4N_2_Dorman_TSF83}.
However, when \pn~is increased to 75 and 100\p, the (111) peak of
the hcp Ni$_3$N phase gets slightly shifted to lower values along
with an additional peak at about 45.5$^{\circ}$. The shift in
(111) peak indicates a further expansion in Ni$_3$N structure
while the additional diffraction peak corresponds to the Ni$_2$N
phase~\cite{PhyB14_Ni2N_film_Nishihara,ASS17_Ni2N_Zhi_yuan}. Such
transformation from the single phased Ni$_3$N to the mixed phase
of Ni$_2$N and Ni$_3$N is also evident from the obtained density
of the films. In table.~\ref{table}, a substantial drop in
X$_{sld}$ can be seen when \pn~is increased from 50 to 75\p,
indicating more nitrogen incorporation in the film  and thus
confirms the formation of N richer Ni$_2$N phase.

\begin{figure} \center
	\vspace{-1mm}
	\includegraphics [scale=0.4] {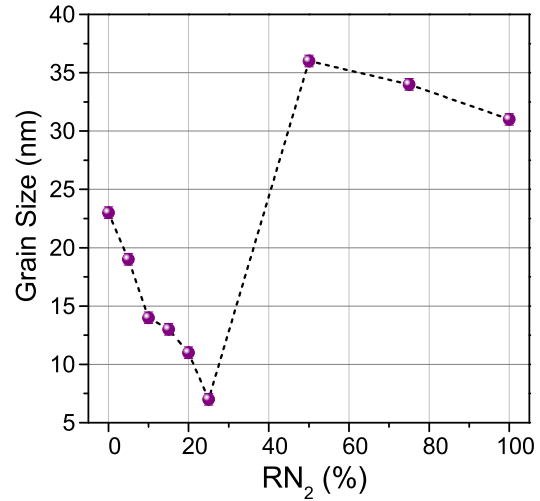}
	\caption{\label{fig:gs} Grain size obtained from the XRD as a
		function of the \pn~for Ni-N films. The error in the grain size
		values are of the size of the symbols. The dashed line is a guide
		to the eye.} \vspace{-1mm}
\end{figure}

The grain size calculated from the XRD peak width using Scherrer
formula is plotted as a function of \pn, is shown in
fig.~\ref{fig:gs}. For the Ni film deposited to \pn~= 0\p, the
grain size obtained is about 23\,nm. On slightly increasing the
\pn~to 5\p, the grain size decreases to about 19\,nm. The grain
size decreases further on increasing \pn~and found to be the
smallest for \pn~= 25\p. Such variation in grain sizes with
\pn~signifies the formation of nanocrystalline grains due to the
formation of the \tnn~phase with the smaller grains. However, the
grain size increases rapidly on further increasing the \pn~at
50\p~which indicates the well-crystalline single phase Ni$_3$N
formation in the films and the grain size increases. However, a
further increase in \pn~would cause the degradation of crystalline
quality and enhance the formation of the Ni$_2$N phase which is
mixed with the Ni$_3$N therefore, the grain size decreased
slightly again. A similar variation in grain sizes as a function
of \pn~has also been previously obtained in a recent
report~\cite{Ni-N_GaN_anode_Li_Vac19}.

\begin{figure} \center
	\vspace{-1mm}
	\includegraphics [scale=0.29] {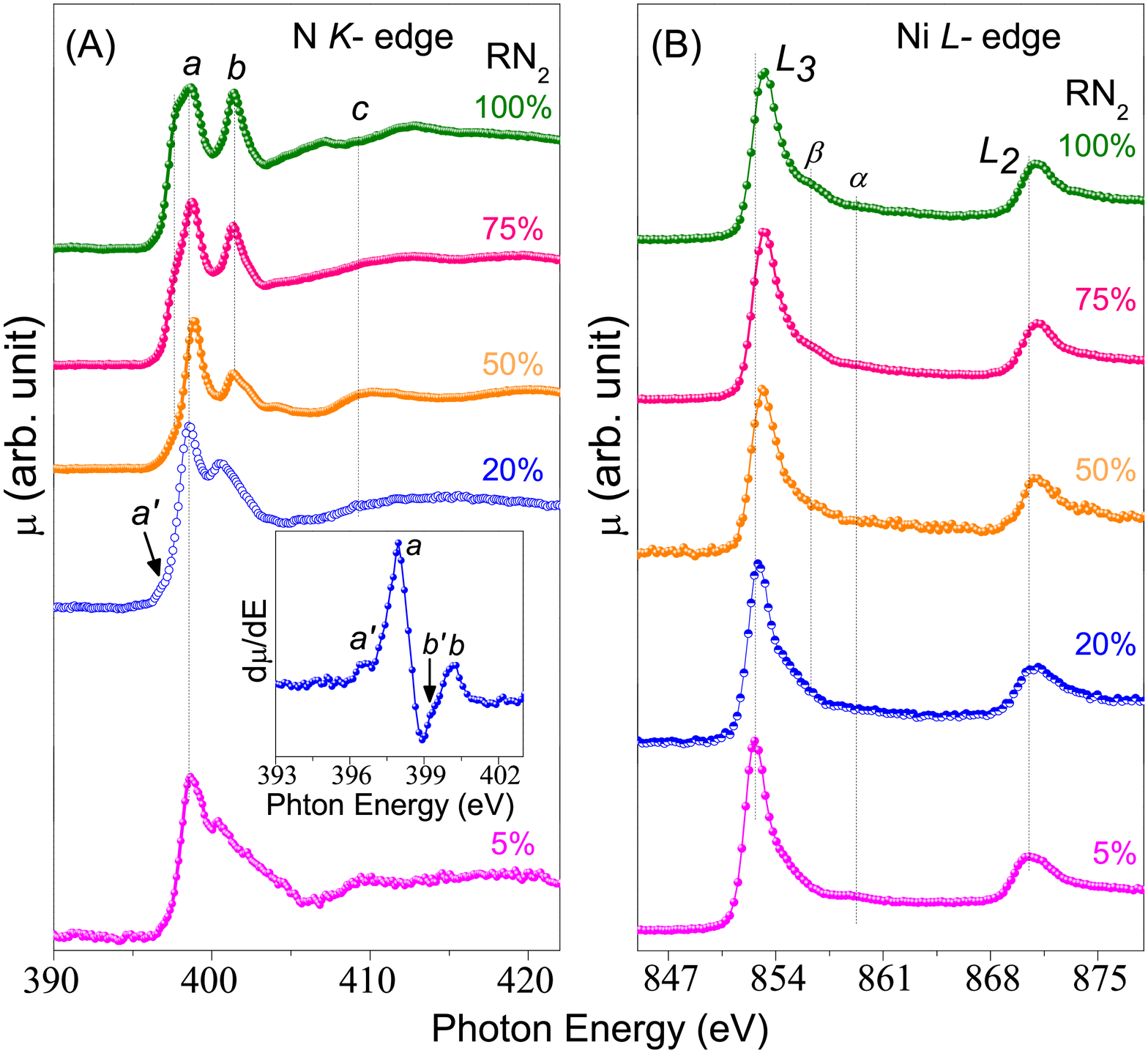}
	\caption{\label{fig:xas} XANES spectra of Ni-N samples deposited
		at \pn~= 5, 20, 50, 75 and 100\p~at N $K-$ edge (A) and Ni $L-$
		edge (B). Inset of (A) shows the derivative of N $K-$ edge XANES
		spectra for Ni-N sample deposited at \pn~= 20\p.} \vspace{-1mm}
\end{figure}

XANES measurements were performed at N $K$ and Ni $L-$ edges of
Ni-N samples, shown in fig.~\ref{fig:xas}. For \pn~= 5\p, a
prominent peak around 398\,eV along with some other features can
be seen in the N $K-$ edge spectra indicating the presence of
incorporated nitrogen in Ni-N sample, can be seen in
fig.~\ref{fig:xas} (A). However, the features get more pronounced
at \pn~= 20\p~due to the enhanced incorporation of nitrogen in
sample. Here, the four feature structures around the main peak at
energies of 397\,eV, 398.5\,eV, 400\,eV, and 401.2\,eV can clearly
be seen in the derivative of N $K-$ edge spectra, assigned as
$a^{\prime}$, $a$, $b^{\prime}$ and $b$ respectively, shown in the
inset of fig.~\ref{fig:xas} (A). Similar behavior has previously
also been probed in the N $K-$ edge spectra for \tfn~thin
films~\cite{JAP_15_K_Ito_XAS}. By calculating the partial
densities of states in \tfn, the origin of different features
present in N $K-$ edge spectra has been well explained in term of
hybridization between Fe 3$d$ and N 2$p$
orbitals~\cite{JAP_15_K_Ito_XAS}. By comparing the N $K-$ edge
spectra obtained for Ni-N sample deposited at \pn~= 20\p~with
those obtained for \tfn~thin film in a work by Ito
$et.~al.$~\cite{JAP_15_K_Ito_XAS}, the formation of \tnn~phase can
also be confirmed which is evident from the XRD results as well.
The origin of feature $a^{\prime}$ can be explained in terms of
$\pi^{\star}$ anti-bonding, and features $a$, $b^{\prime}$ and $b$
are explained by $\sigma^{\star}$ anti-bonding states arises due
to dipole transition from the N 1$s$ core-level to the hybridized
states of Ni 3$d$ and N 2$p$. For \pn~$\geq$ 50\p, a noticeable
change in N $K-$ edge XANES spectra appears which may be due to
the phase transformation from \tnn~phase to other Ni-N phases. At
\pn~= 50, 75 and 100\p, now the features $a$, $b$ may be
attributed to the $\pi^{\star}$ anti-bonding formed by
hybridization between Ni 3$d_{xy}$, 3$d_{yz}$, 3$d_{zx}$, and N
2$p$ orbitals, and the $\sigma^{\star}$ anti-bonding due to Ni
3d$_{z^2-r^2}$ and N 2$p$ hybridization, respectively. While the
feature $c$ and above may attributed to transition from N 1$s$ to
hybridized states of N 2$p$ and Ni 4$sp$ states.

The Ni $L-$ edge spectra shows two main peaks at about 852.5 and
870\,eV correspond to $L_3$ and $L_2$ edges arises due to
well-known spin-orbit interaction, shown in fig.~\ref{fig:xas}
(B). A weak satellite feature around 858.5\,eV assigned as
$\alpha$ can also be seen which corresponds to hybridization
between valence $d$ and unoccupied $sp$ states. A gradual shift in
the peak position of $L_3$ edge centroid to higher energy side can
be seen with increasing \pn~compared to 5\p~spectra. Such behavior
again signifies the increased oxidation state of Ni in different
Ni-N phases. In addition, a small feature appears at about 2\,eV
above the $L_3$ edge assigned as $\beta$ in fig.~\ref{fig:xas} (B)
which is prominently present in \pn $\geq$ 75\p. This feature
$\beta$ has previously been obtained for Ni +2 state species and
arises due to strong interactions between core holes and 3$d$
orbitals of Ni~\cite{Ni_XAS_+2state}. In the present study also,
we have seen that at \pn $\geq$ 75\p, a mixed phase of Ni$_2$N and
Ni$_3$N is present in the sample and the oxidation state of Ni is
1.5 in Ni$_2$N phase~\cite{ASS17_Ni2N_Zhi_yuan}. Therefore, the
presence of this feature $\beta$ and upshift of the $L_3$ edge
centroid directly reflects nearly +2 chemical valency of Ni in \pn
$\geq$ 75\p~and supports the formation of Ni$_2$N phase as well.

Hence, from XANES results also, it is clear that with the
successive increment in the \pn, different Ni-N phases get formed
which is well-consistent with our XRD results.

\begin{table*} \footnotesize \center \vspace{-5mm}
	\caption{\label{table} Parameters, \pn, roughness obtained from
		XRR, phase identified by XRD, scattering length density
		(X$_{sld}$) measured from XRR, theoretical X$_{sld}$, nuclear sld
		(N$_{sld}$) obtained from PNR, theoretical N$_{sld}$ and magnetic
		moment ($M_s$) of Ni-N thin film samples.}
	\begin{tabular}{cccccccc}\hline
		
		\pn &Roughness&Phase(s) identified               &X$_{sld}$               &X$_{sld}$          &N$_{sld}$      &N$_{sld}$   &$M_s$\\
		&XRR      &XRD                               &Exp.                    &Theo.               &Exp.           &Theo.        &Exp.\\
		(\p)&\AA      &
		&\AA$^{-2}$&\AA$^{-2}$&\AA$^{-2}$&\AA$^{-2}$
		&\muB/Ni\\\hline\hline

		0&4&Ni                  &6.3$\times$10$^{-5}$ &6.44$\times$10$^{-5}$  &9.1$\times$10$^{-6}$           &9.41$\times$10$^{-6}$&0.45\\
		5&5&Ni(N)               &-                    &-                      &9.0$\times$10$^{-6}$           &-                    &0.18\\
		12&5&Ni$_{4+x}$N$_{1-x}$&5.9$\times$10$^{-5}$ &-                      &8.9$\times$10$^{-6}$           &-                    &0\\
		20&-&\tnn               & -                   &5.77$\times$10$^{-5}$  &8.86$\times$10$^{-6}$          &9.66$\times$10$^{-6}$&0\\
		25&5&\tnn+Ni$_3$N       &5.8$\times$10$^{-5}$ &-                      &9.1$\times$10$^{-6}$           &-                    &0\\
		50&6&Ni$_3$N            &5.6$\times$10$^{-5}$ &5.7$\times$10$^{-5}$   &9.2$\times$10$^{-6}$           &1.02$\times$10$^{-5}$&0\\
		75&8&Ni$_3$N+Ni$_2$N    &4.8$\times$10$^{-5}$ & -                     &9.5$\times$10$^{-6}$           &-                    &0\\
		100&7&Ni$_3$N+Ni$_2$N   &5.2$\times$10$^{-5}$
		&5.52$\times$10$^{-5}$$^{\star}$
		&7.6$\times$10$^{-6}$&1.03$\times$10$^{-5}$$^{\star}$&0\\\hline
		
	\end{tabular}
	
	\flushleft{$^{\star}$For the pure Ni$_2$N phase.}\\
\end{table*}

\subsection{Magnetization Measurements}
\begin{figure}\center
	\vspace{-1mm}
	\includegraphics [scale=0.4] {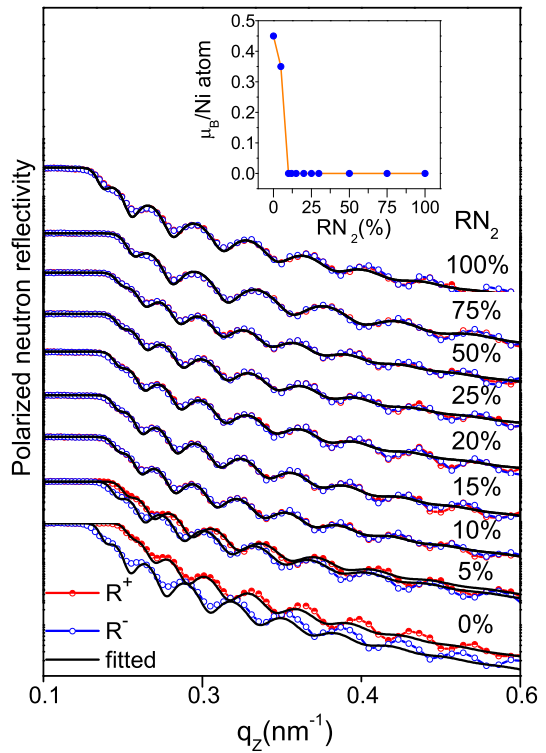}
	\caption{\label{fig:pnr_all} PNR patterns taken at 300\,K of Ni-N
		thin films deposited at different values of \pn. Inset shows that
		variation of $M_s$ with \pn.} \vspace{-1mm}
\end{figure}

We performed magnetization measurements along out-of-plane
direction in the Ni-N samples using SQUID-VSM and magneto-optical
Kerr effect (not shown here) measurements. We found the absence of
out-of-plane magnetization component in the samples. Therefore, to
probe the in-plane magnetization in the Ni-N samples, PNR
measurements were carried out as PNR is ideally favorable to
determine the nuclear and the averaged in-plane magnetization
depth profile of the film. The PNR patterns of Ni-N films are
shown in fig.~\ref{fig:pnr_all}. We can see that the splitting
between spin-up (R$^+$) and down (R$^-$) reflectivities in PNR
patterns are only visible for \pn~= 0 and 5\p~while disappears
afterwards. Since, it is known that the splitting between R$^+$
and R$^-$ reflectivity in the PNR pattern clearly indicates the
ferromagnetic state of the sample~\cite{PNR92_Blundell}.
Therefore, such behavior in PNR patterns directly signifies the
ferromagnetic ordering for \pn~= 0 and 5\p~while a non-magnetic
state for \pn~$>$ 5\p~in Ni-N samples. The pure Ni sample
(\pn~=0\p) is fitted (using SimulReflec
programme~\cite{SimulReflec}) considering the magnetic moment
($M_s$) of 0.45\,\muB/Ni while it reduces to 0.18\,\muB/Ni for
\pn~= 5\p. From the PNR data, we can say that ferromagnetism
retains only up to \pn~= 5\p, and disappears afterward as shown in
the inset of fig.~\ref{fig:pnr_all}. However, the Ni$_3$N and
Ni$_2$N phases have previously been reported to show paramagnetic
behavior~\cite{JAC09_Ni3N_film_Vempaire,JAP09_Ni3N_film_Vempaire,ASS09_Popovic_Ni3N}.
Therefore, the non-magnetic behavior of Ni-N films deposited at
\pn~= 25 to 100\p~is expected. On the other hand, it has been
theoretically reported that the \tnn~phase posses a small magnetic
moment 0.32\,\muB/Ni with Curie temperature of $\approx$
121\,K~\cite{JPCM_Markus16}. In the present study, we found no
ferromagnetism in \tnn~films deposited at \pn~$\leq$ 20\pat~at
room temperature, as confirmed by our PNR measurements. Contrary
to the present observation, other experimental works claimed a
ferromagnetic behavior with Curie temperature $\approx$ 500\,K in
the \tnn~films. Although, it is to be noted here that those Ni-N
films were prepared at high \Ts~of about
455-625\,K~\cite{Mag_NiN_Linnik2013,Ni4N_high_Ts_Shalayev2014}.
Therefore, the film may contain a dominant Ni phase that may be
responsible for the presence of ferromagnetism at room
temperature~\cite{Mag_NiN_Linnik2013,Ni4N_high_Ts_Shalayev2014}.

\section{Conclusion}
We have successfully synthesized the different phases of the Ni-N
system by reactive magnetron sputtering process. The deposition
rates and densities of Ni-N films decreases with increasing \pn,
indicates the target poisoning behavior and gradual incorporation
of N in Ni-N thin films, respectively. Evolution of different Ni-N
phases; Ni(N)$\longrightarrow$\tnn$\longrightarrow$
Ni$_3$N$\longrightarrow$Ni$_2$N phases obtained with successively
increasing \pn, confirmed by XRD results. The transformation into
different Ni-N phases is also reflected in their corresponding
grain sizes. The formation of different Ni-N phases with varying
\pn~is also further supported by XANES results. PNR measurements
reveal the presence of ferromagnetism only in pure Ni (\pn~= 0\p)
and in Ni-N sample deposited at a very smaller value of \pn~= 5\p,
while the Ni-N samples become non-magnetic for \pn~$>$ 5\p.

\section*{Acknowledgments}
On of the author (NP) is thankful to Council of Scientific
Industrial Research (CSIR) for senior research fellowship (SRF).
Authors thank the Department of Science and Technology, India
(SR/NM/Z-07/2015) for the financial support and Jawaharlal Nehru
Centre for Advanced Scientific Research (JNCASR) for managing the
project. A part of this work was performed at AMOR, Swiss
Spallation Neutron Source, Paul Scherrer Institute, Villigen,
Switzerland. D. M. Phase is acknowledged for the BL-01
measurements. We acknowledge V. R. Reddy and A. Gome for XRR
measurements and Layanta Behera for the help provided in various
experiments.


\begin{thebibliography}{74}%
	\makeatletter
	\providecommand \@ifxundefined [1]{%
		\@ifx{#1\undefined}
	}%
	\providecommand \@ifnum [1]{%
		\ifnum #1\expandafter \@firstoftwo
		\else \expandafter \@secondoftwo
		\fi
	}%
	\providecommand \@ifx [1]{%
		\ifx #1\expandafter \@firstoftwo
		\else \expandafter \@secondoftwo
		\fi
	}%
	\providecommand \natexlab [1]{#1}%
	\providecommand \enquote  [1]{``#1''}%
	\providecommand \bibnamefont  [1]{#1}%
	\providecommand \bibfnamefont [1]{#1}%
	\providecommand \citenamefont [1]{#1}%
	\providecommand \href@noop [0]{\@secondoftwo}%
	\providecommand \href [0]{\begingroup \@sanitize@url \@href}%
	\providecommand \@href[1]{\@@startlink{#1}\@@href}%
	\providecommand \@@href[1]{\endgroup#1\@@endlink}%
	\providecommand \@sanitize@url [0]{\catcode `\\12\catcode `\$12\catcode
		`\&12\catcode `\#12\catcode `\^12\catcode `\_12\catcode `\%12\relax}%
	\providecommand \@@startlink[1]{}%
	\providecommand \@@endlink[0]{}%
	\providecommand \url  [0]{\begingroup\@sanitize@url \@url }%
	\providecommand \@url [1]{\endgroup\@href {#1}{\urlprefix }}%
	\providecommand \urlprefix  [0]{URL }%
	\providecommand \Eprint [0]{\href }%
	\providecommand \doibase [0]{http://dx.doi.org/}%
	\providecommand \selectlanguage [0]{\@gobble}%
	\providecommand \bibinfo  [0]{\@secondoftwo}%
	\providecommand \bibfield  [0]{\@secondoftwo}%
	\providecommand \translation [1]{[#1]}%
	\providecommand \BibitemOpen [0]{}%
	\providecommand \bibitemStop [0]{}%
	\providecommand \bibitemNoStop [0]{.\EOS\space}%
	\providecommand \EOS [0]{\spacefactor3000\relax}%
	\providecommand \BibitemShut  [1]{\csname bibitem#1\endcsname}%
	\let\auto@bib@innerbib\@empty
	\bibitem [{\citenamefont {Sun}\ \emph {et~al.}(2019)\citenamefont {Sun},
		\citenamefont {Bartel}, \citenamefont {Arca}, \citenamefont {Bauers},
		\citenamefont {Matthews}, \citenamefont {Orva{\~n}anos}, \citenamefont
		{Chen}, \citenamefont {Toney}, \citenamefont {Schelhas}, \citenamefont
		{Tumas} \emph {et~al.}}]{TMN_NatMat19_Sun_Map}%
	\BibitemOpen
	\bibfield  {author} {\bibinfo {author} {\bibfnamefont {W.}~\bibnamefont
			{Sun}}, \bibinfo {author} {\bibfnamefont {C.~J.}\ \bibnamefont {Bartel}},
		\bibinfo {author} {\bibfnamefont {E.}~\bibnamefont {Arca}}, \bibinfo {author}
		{\bibfnamefont {S.~R.}\ \bibnamefont {Bauers}}, \bibinfo {author}
		{\bibfnamefont {B.}~\bibnamefont {Matthews}}, \bibinfo {author}
		{\bibfnamefont {B.}~\bibnamefont {Orva{\~n}anos}}, \bibinfo {author}
		{\bibfnamefont {B.-R.}\ \bibnamefont {Chen}}, \bibinfo {author}
		{\bibfnamefont {M.~F.}\ \bibnamefont {Toney}}, \bibinfo {author}
		{\bibfnamefont {L.~T.}\ \bibnamefont {Schelhas}}, \bibinfo {author}
		{\bibfnamefont {W.}~\bibnamefont {Tumas}},  \emph {et~al.},\ }\href@noop {}
	{\bibfield  {journal} {\bibinfo  {journal} {Nature materials}\ ,\ \bibinfo
			{pages} {1}} (\bibinfo {year} {2019})}\BibitemShut {NoStop}%
	\bibitem [{\citenamefont {Khazaei}\ \emph {et~al.}(2013)\citenamefont
		{Khazaei}, \citenamefont {Arai}, \citenamefont {Sasaki}, \citenamefont
		{Chung}, \citenamefont {Venkataramanan}, \citenamefont {Estili},
		\citenamefont {Sakka},\ and\ \citenamefont {Kawazoe}}]{Khazaei_2013_TMN}%
	\BibitemOpen
	\bibfield  {author} {\bibinfo {author} {\bibfnamefont {M.}~\bibnamefont
			{Khazaei}}, \bibinfo {author} {\bibfnamefont {M.}~\bibnamefont {Arai}},
		\bibinfo {author} {\bibfnamefont {T.}~\bibnamefont {Sasaki}}, \bibinfo
		{author} {\bibfnamefont {C.-Y.}\ \bibnamefont {Chung}}, \bibinfo {author}
		{\bibfnamefont {N.~S.}\ \bibnamefont {Venkataramanan}}, \bibinfo {author}
		{\bibfnamefont {M.}~\bibnamefont {Estili}}, \bibinfo {author} {\bibfnamefont
			{Y.}~\bibnamefont {Sakka}}, \ and\ \bibinfo {author} {\bibfnamefont
			{Y.}~\bibnamefont {Kawazoe}},\ }\href@noop {} {\bibfield  {journal} {\bibinfo
			{journal} {Advanced Functional Materials}\ }\textbf {\bibinfo {volume}
			{23}},\ \bibinfo {pages} {2185} (\bibinfo {year} {2013})}\BibitemShut
	{NoStop}%
	\bibitem [{\citenamefont {Zhong}\ \emph {et~al.}(2016)\citenamefont {Zhong},
		\citenamefont {Xia}, \citenamefont {Shi}, \citenamefont {Zhan}, \citenamefont
		{Tu},\ and\ \citenamefont {Fan}}]{TMN_Zhong16}%
	\BibitemOpen
	\bibfield  {author} {\bibinfo {author} {\bibfnamefont {Y.}~\bibnamefont
			{Zhong}}, \bibinfo {author} {\bibfnamefont {X.}~\bibnamefont {Xia}}, \bibinfo
		{author} {\bibfnamefont {F.}~\bibnamefont {Shi}}, \bibinfo {author}
		{\bibfnamefont {J.}~\bibnamefont {Zhan}}, \bibinfo {author} {\bibfnamefont
			{J.}~\bibnamefont {Tu}}, \ and\ \bibinfo {author} {\bibfnamefont {H.~J.}\
			\bibnamefont {Fan}},\ }\href@noop {} {\bibfield  {journal} {\bibinfo
			{journal} {Advanced science}\ }\textbf {\bibinfo {volume} {3}},\ \bibinfo
		{pages} {1500286} (\bibinfo {year} {2016})}\BibitemShut {NoStop}%
	\bibitem [{\citenamefont {Oyama}(1996)}]{TMN_Oyama_96}%
	\BibitemOpen
	\bibfield  {author} {\bibinfo {author} {\bibfnamefont {S.~T.}\ \bibnamefont
			{Oyama}},\ }\href@noop {} {\emph {\bibinfo {title} {Introduction to the
				chemistry of transition metal carbides and nitrides}}}\ (\bibinfo
	{publisher} {Springer, Dordrecht},\ \bibinfo {year} {1996})\ pp.\ \bibinfo
	{pages} {1--27}\BibitemShut {NoStop}%
	\bibitem [{\citenamefont {Williams}(1997)}]{TMN_Williams_97}%
	\BibitemOpen
	\bibfield  {author} {\bibinfo {author} {\bibfnamefont {W.~S.}\ \bibnamefont
			{Williams}},\ }\href@noop {} {\bibfield  {journal} {\bibinfo  {journal}
			{Jom}\ }\textbf {\bibinfo {volume} {49}},\ \bibinfo {pages} {38} (\bibinfo
		{year} {1997})}\BibitemShut {NoStop}%
	\bibitem [{\citenamefont {Coey}\ and\ \citenamefont
		{Smith}(1999)}]{JMMM90_Mag_Nit_Coey}%
	\BibitemOpen
	\bibfield  {author} {\bibinfo {author} {\bibfnamefont {J.}~\bibnamefont
			{Coey}}\ and\ \bibinfo {author} {\bibfnamefont {P.}~\bibnamefont {Smith}},\
	}\href@noop {} {\bibfield  {journal} {\bibinfo  {journal} {Journal of
				magnetism and magnetic materials}\ }\textbf {\bibinfo {volume} {200}},\
		\bibinfo {pages} {405} (\bibinfo {year} {1999})}\BibitemShut {NoStop}%
	\bibitem [{\citenamefont {Ningthoujam}\ and\ \citenamefont
		{Gajbhiye}(2015)}]{TMN_Ningthoujam_Gajbhiye_15}%
	\BibitemOpen
	\bibfield  {author} {\bibinfo {author} {\bibfnamefont {R.}~\bibnamefont
			{Ningthoujam}}\ and\ \bibinfo {author} {\bibfnamefont {N.}~\bibnamefont
			{Gajbhiye}},\ }\href@noop {} {\bibfield  {journal} {\bibinfo  {journal}
			{Progress in Materials Science}\ }\textbf {\bibinfo {volume} {70}},\ \bibinfo
		{pages} {50} (\bibinfo {year} {2015})}\BibitemShut {NoStop}%
	\bibitem [{\citenamefont {Gupta}(2020)}]{MG_book_Fe-N}%
	\BibitemOpen
	\bibfield  {author} {\bibinfo {author} {\bibfnamefont {M.}~\bibnamefont
			{Gupta}},\ }\href@noop {} {\emph {\bibinfo {title} {Synthesis, Stability and
				Self-Diffusion in Iron Nitride Thin Films: A Review}}}\ (\bibinfo
	{publisher} {to be published in Springer},\ \bibinfo {year}
	{2020})\BibitemShut {NoStop}%
	\bibitem [{\citenamefont {Yang}\ \emph {et~al.}(2018)\citenamefont {Yang},
		\citenamefont {Tao}, \citenamefont {Jiang}, \citenamefont {Chen},
		\citenamefont {Tang}, \citenamefont {Yan},\ and\ \citenamefont
		{Han}}]{PRB18_Theo_Fe4N/Co4N/MgO}%
	\BibitemOpen
	\bibfield  {author} {\bibinfo {author} {\bibfnamefont {B.}~\bibnamefont
			{Yang}}, \bibinfo {author} {\bibfnamefont {L.}~\bibnamefont {Tao}}, \bibinfo
		{author} {\bibfnamefont {L.}~\bibnamefont {Jiang}}, \bibinfo {author}
		{\bibfnamefont {W.}~\bibnamefont {Chen}}, \bibinfo {author} {\bibfnamefont
			{P.}~\bibnamefont {Tang}}, \bibinfo {author} {\bibfnamefont {Y.}~\bibnamefont
			{Yan}}, \ and\ \bibinfo {author} {\bibfnamefont {X.}~\bibnamefont {Han}},\
	}\href@noop {} {\bibfield  {journal} {\bibinfo  {journal} {Physical Review
				Applied}\ }\textbf {\bibinfo {volume} {9}},\ \bibinfo {pages} {054019}
		(\bibinfo {year} {2018})}\BibitemShut {NoStop}%
	\bibitem [{\citenamefont {Bhattacharyya}(2015)}]{FeN_Review_Bhattacharyya15}%
	\BibitemOpen
	\bibfield  {author} {\bibinfo {author} {\bibfnamefont {S.}~\bibnamefont
			{Bhattacharyya}},\ }\href@noop {} {\bibfield  {journal} {\bibinfo  {journal}
			{The Journal of Physical Chemistry C}\ }\textbf {\bibinfo {volume} {119}},\
		\bibinfo {pages} {1601} (\bibinfo {year} {2015})}\BibitemShut {NoStop}%
	\bibitem [{\citenamefont {Hui}\ \emph {et~al.}(2018)\citenamefont {Hui},
		\citenamefont {Xie}, \citenamefont {Li},\ and\ \citenamefont
		{Chen}}]{FeX_N_B_JMMM18_Hui18}%
	\BibitemOpen
	\bibfield  {author} {\bibinfo {author} {\bibfnamefont {L.}~\bibnamefont
			{Hui}}, \bibinfo {author} {\bibfnamefont {Z.}~\bibnamefont {Xie}}, \bibinfo
		{author} {\bibfnamefont {C.}~\bibnamefont {Li}}, \ and\ \bibinfo {author}
		{\bibfnamefont {Z.-Q.}\ \bibnamefont {Chen}},\ }\href@noop {} {\bibfield
		{journal} {\bibinfo  {journal} {Journal of Magnetism and Magnetic Materials}\
		}\textbf {\bibinfo {volume} {451}},\ \bibinfo {pages} {761} (\bibinfo {year}
		{2018})}\BibitemShut {NoStop}%
	\bibitem [{\citenamefont {Blanc{\'a}}\ \emph {et~al.}(2009)\citenamefont
		{Blanc{\'a}}, \citenamefont {Desimoni}, \citenamefont {Christensen},
		\citenamefont {Emmerich},\ and\ \citenamefont
		{Cottenier}}]{Fe4N_Blanca09_Review}%
	\BibitemOpen
	\bibfield  {author} {\bibinfo {author} {\bibfnamefont {E.~L. P.~y.}\
			\bibnamefont {Blanc{\'a}}}, \bibinfo {author} {\bibfnamefont
			{J.}~\bibnamefont {Desimoni}}, \bibinfo {author} {\bibfnamefont {N.~E.}\
			\bibnamefont {Christensen}}, \bibinfo {author} {\bibfnamefont
			{H.}~\bibnamefont {Emmerich}}, \ and\ \bibinfo {author} {\bibfnamefont
			{S.}~\bibnamefont {Cottenier}},\ }\href@noop {} {\bibfield  {journal}
		{\bibinfo  {journal} {physica status solidi (b)}\ }\textbf {\bibinfo {volume}
			{246}},\ \bibinfo {pages} {909} (\bibinfo {year} {2009})}\BibitemShut
	{NoStop}%
	\bibitem [{\citenamefont {Meinert}(2016)}]{JPCM_Markus16}%
	\BibitemOpen
	\bibfield  {author} {\bibinfo {author} {\bibfnamefont {M.}~\bibnamefont
			{Meinert}},\ }\href@noop {} {\bibfield  {journal} {\bibinfo  {journal}
			{Journal of Physics: Condensed Matter}\ }\textbf {\bibinfo {volume} {28}},\
		\bibinfo {pages} {056006} (\bibinfo {year} {2016})}\BibitemShut {NoStop}%
	\bibitem [{\citenamefont {Imai}\ \emph {et~al.}(2010)\citenamefont {Imai},
		\citenamefont {Takahashi},\ and\ \citenamefont {Kumagai}}]{JMMM10_Imai}%
	\BibitemOpen
	\bibfield  {author} {\bibinfo {author} {\bibfnamefont {Y.}~\bibnamefont
			{Imai}}, \bibinfo {author} {\bibfnamefont {Y.}~\bibnamefont {Takahashi}}, \
		and\ \bibinfo {author} {\bibfnamefont {T.}~\bibnamefont {Kumagai}},\ }\href
	{\doibase http://dx.doi.org/10.1016/j.jmmm.2010.04.004} {\bibfield  {journal}
		{\bibinfo  {journal} {Journal of Magnetism and Magnetic Materials}\ }\textbf
		{\bibinfo {volume} {322}},\ \bibinfo {pages} {2665 } (\bibinfo {year}
		{2010})}\BibitemShut {NoStop}%
	\bibitem [{\citenamefont {Imai}\ \emph {et~al.}(2014)\citenamefont {Imai},
		\citenamefont {Sohma},\ and\ \citenamefont {Suemasu}}]{JAC14_Imai}%
	\BibitemOpen
	\bibfield  {author} {\bibinfo {author} {\bibfnamefont {Y.}~\bibnamefont
			{Imai}}, \bibinfo {author} {\bibfnamefont {M.}~\bibnamefont {Sohma}}, \ and\
		\bibinfo {author} {\bibfnamefont {T.}~\bibnamefont {Suemasu}},\ }\href
	{\doibase http://dx.doi.org/10.1016/j.jallcom.2014.04.171} {\bibfield
		{journal} {\bibinfo  {journal} {Journal of Alloys and Compounds}\ }\textbf
		{\bibinfo {volume} {611}},\ \bibinfo {pages} {440 } (\bibinfo {year}
		{2014})}\BibitemShut {NoStop}%
	\bibitem [{\citenamefont {Metzger}\ \emph {et~al.}(1994)\citenamefont
		{Metzger}, \citenamefont {Bao},\ and\ \citenamefont
		{Carbucicchio}}]{Fe16N2_Metzger94}%
	\BibitemOpen
	\bibfield  {author} {\bibinfo {author} {\bibfnamefont {R.~M.}\ \bibnamefont
			{Metzger}}, \bibinfo {author} {\bibfnamefont {X.}~\bibnamefont {Bao}}, \ and\
		\bibinfo {author} {\bibfnamefont {M.}~\bibnamefont {Carbucicchio}},\
	}\href@noop {} {\bibfield  {journal} {\bibinfo  {journal} {Journal of Applied
				Physics}\ }\textbf {\bibinfo {volume} {76}},\ \bibinfo {pages} {6626}
		(\bibinfo {year} {1994})}\BibitemShut {NoStop}%
	\bibitem [{\citenamefont {Matar}(1992)}]{Fe8N_Matar92}%
	\BibitemOpen
	\bibfield  {author} {\bibinfo {author} {\bibfnamefont {S.}~\bibnamefont
			{Matar}},\ }\href@noop {} {\bibfield  {journal} {\bibinfo  {journal}
			{Zeitschrift f{\"u}r Physik B Condensed Matter}\ }\textbf {\bibinfo {volume}
			{87}},\ \bibinfo {pages} {91} (\bibinfo {year} {1992})}\BibitemShut {NoStop}%
	\bibitem [{\citenamefont {Leineweber}\ \emph {et~al.}(1999)\citenamefont
		{Leineweber}, \citenamefont {Jacobs}, \citenamefont {Hüning}, \citenamefont
		{Lueken}, \citenamefont {Schilder},\ and\ \citenamefont
		{Kockelmann}}]{Fe3N_Leineweber_JAC99}%
	\BibitemOpen
	\bibfield  {author} {\bibinfo {author} {\bibfnamefont {A.}~\bibnamefont
			{Leineweber}}, \bibinfo {author} {\bibfnamefont {H.}~\bibnamefont {Jacobs}},
		\bibinfo {author} {\bibfnamefont {F.}~\bibnamefont {Hüning}}, \bibinfo
		{author} {\bibfnamefont {H.}~\bibnamefont {Lueken}}, \bibinfo {author}
		{\bibfnamefont {H.}~\bibnamefont {Schilder}}, \ and\ \bibinfo {author}
		{\bibfnamefont {W.}~\bibnamefont {Kockelmann}},\ }\href {\doibase
		https://doi.org/10.1016/S0925-8388(99)00150-4} {\bibfield  {journal}
		{\bibinfo  {journal} {Journal of Alloys and Compounds}\ }\textbf {\bibinfo
			{volume} {288}},\ \bibinfo {pages} {79 } (\bibinfo {year}
		{1999})}\BibitemShut {NoStop}%
	\bibitem [{\citenamefont {Rechenbach}\ and\ \citenamefont
		{Jacobs}(1996)}]{Fe2N_JAC18}%
	\BibitemOpen
	\bibfield  {author} {\bibinfo {author} {\bibfnamefont {D.}~\bibnamefont
			{Rechenbach}}\ and\ \bibinfo {author} {\bibfnamefont {H.}~\bibnamefont
			{Jacobs}},\ }\href {\doibase https://doi.org/10.1016/0925-8388(95)02097-7}
	{\bibfield  {journal} {\bibinfo  {journal} {Journal of Alloys and Compounds}\
		}\textbf {\bibinfo {volume} {235}},\ \bibinfo {pages} {15 } (\bibinfo {year}
		{1996})}\BibitemShut {NoStop}%
	\bibitem [{\citenamefont {Zhao}\ \emph {et~al.}(2015)\citenamefont {Zhao},
		\citenamefont {Bao}, \citenamefont {Duan}, \citenamefont {Tian},
		\citenamefont {Liu},\ and\ \citenamefont {Cui}}]{FeN_Zhao_RSC15}%
	\BibitemOpen
	\bibfield  {author} {\bibinfo {author} {\bibfnamefont {Z.}~\bibnamefont
			{Zhao}}, \bibinfo {author} {\bibfnamefont {K.}~\bibnamefont {Bao}}, \bibinfo
		{author} {\bibfnamefont {D.}~\bibnamefont {Duan}}, \bibinfo {author}
		{\bibfnamefont {F.}~\bibnamefont {Tian}}, \bibinfo {author} {\bibfnamefont
			{B.}~\bibnamefont {Liu}}, \ and\ \bibinfo {author} {\bibfnamefont
			{T.}~\bibnamefont {Cui}},\ }\href@noop {} {\bibfield  {journal} {\bibinfo
			{journal} {Rsc Advances}\ }\textbf {\bibinfo {volume} {5}},\ \bibinfo {pages}
		{31270} (\bibinfo {year} {2015})}\BibitemShut {NoStop}%
	\bibitem [{\citenamefont {Niwa}\ \emph {et~al.}(2017)\citenamefont {Niwa},
		\citenamefont {Terabe}, \citenamefont {Kato}, \citenamefont {Takayama},
		\citenamefont {Kato}, \citenamefont {Soda},\ and\ \citenamefont
		{Hasegawa}}]{FeNx_CoNx_Niwa_ACS17}%
	\BibitemOpen
	\bibfield  {author} {\bibinfo {author} {\bibfnamefont {K.}~\bibnamefont
			{Niwa}}, \bibinfo {author} {\bibfnamefont {T.}~\bibnamefont {Terabe}},
		\bibinfo {author} {\bibfnamefont {D.}~\bibnamefont {Kato}}, \bibinfo {author}
		{\bibfnamefont {S.}~\bibnamefont {Takayama}}, \bibinfo {author}
		{\bibfnamefont {M.}~\bibnamefont {Kato}}, \bibinfo {author} {\bibfnamefont
			{K.}~\bibnamefont {Soda}}, \ and\ \bibinfo {author} {\bibfnamefont
			{M.}~\bibnamefont {Hasegawa}},\ }\href@noop {} {\bibfield  {journal}
		{\bibinfo  {journal} {Inorganic chemistry}\ }\textbf {\bibinfo {volume}
			{56}},\ \bibinfo {pages} {6410} (\bibinfo {year} {2017})}\BibitemShut
	{NoStop}%
	\bibitem [{\citenamefont {Bykov}\ \emph {et~al.}(2018)\citenamefont {Bykov},
		\citenamefont {Bykova}, \citenamefont {Aprilis}, \citenamefont {Glazyrin},
		\citenamefont {Koemets}, \citenamefont {Chuvashova}, \citenamefont {Kupenko},
		\citenamefont {McCammon}, \citenamefont {Mezouar}, \citenamefont {Prakapenka}
		\emph {et~al.}}]{FeNx_Bykov_Nat18}%
	\BibitemOpen
	\bibfield  {author} {\bibinfo {author} {\bibfnamefont {M.}~\bibnamefont
			{Bykov}}, \bibinfo {author} {\bibfnamefont {E.}~\bibnamefont {Bykova}},
		\bibinfo {author} {\bibfnamefont {G.}~\bibnamefont {Aprilis}}, \bibinfo
		{author} {\bibfnamefont {K.}~\bibnamefont {Glazyrin}}, \bibinfo {author}
		{\bibfnamefont {E.}~\bibnamefont {Koemets}}, \bibinfo {author} {\bibfnamefont
			{I.}~\bibnamefont {Chuvashova}}, \bibinfo {author} {\bibfnamefont
			{I.}~\bibnamefont {Kupenko}}, \bibinfo {author} {\bibfnamefont
			{C.}~\bibnamefont {McCammon}}, \bibinfo {author} {\bibfnamefont
			{M.}~\bibnamefont {Mezouar}}, \bibinfo {author} {\bibfnamefont
			{V.}~\bibnamefont {Prakapenka}},  \emph {et~al.},\ }\href@noop {} {\bibfield
		{journal} {\bibinfo  {journal} {Nature communications}\ }\textbf {\bibinfo
			{volume} {9}},\ \bibinfo {pages} {2756} (\bibinfo {year} {2018})}\BibitemShut
	{NoStop}%
	\bibitem [{\citenamefont {Wu}\ \emph {et~al.}(2018)\citenamefont {Wu},
		\citenamefont {Tian}, \citenamefont {Wan}, \citenamefont {Liu}, \citenamefont
		{Gong}, \citenamefont {Chen}, \citenamefont {Shen}, \citenamefont {Yao},
		\citenamefont {Gou},\ and\ \citenamefont {Gao}}]{FeNx_Lailei_ACS18}%
	\BibitemOpen
	\bibfield  {author} {\bibinfo {author} {\bibfnamefont {L.}~\bibnamefont
			{Wu}}, \bibinfo {author} {\bibfnamefont {R.}~\bibnamefont {Tian}}, \bibinfo
		{author} {\bibfnamefont {B.}~\bibnamefont {Wan}}, \bibinfo {author}
		{\bibfnamefont {H.}~\bibnamefont {Liu}}, \bibinfo {author} {\bibfnamefont
			{N.}~\bibnamefont {Gong}}, \bibinfo {author} {\bibfnamefont {P.}~\bibnamefont
			{Chen}}, \bibinfo {author} {\bibfnamefont {T.}~\bibnamefont {Shen}}, \bibinfo
		{author} {\bibfnamefont {Y.}~\bibnamefont {Yao}}, \bibinfo {author}
		{\bibfnamefont {H.}~\bibnamefont {Gou}}, \ and\ \bibinfo {author}
		{\bibfnamefont {F.}~\bibnamefont {Gao}},\ }\href@noop {} {\bibfield
		{journal} {\bibinfo  {journal} {Chemistry of Materials}\ }\textbf {\bibinfo
			{volume} {30}},\ \bibinfo {pages} {8476} (\bibinfo {year}
		{2018})}\BibitemShut {NoStop}%
	\bibitem [{\citenamefont {Pandey}\ \emph
		{et~al.}(2019{\natexlab{a}})\citenamefont {Pandey}, \citenamefont {Gupta},
		\citenamefont {Gupta}, \citenamefont {Amir},\ and\ \citenamefont
		{Stahn}}]{APA19_Co4N_NP}%
	\BibitemOpen
	\bibfield  {author} {\bibinfo {author} {\bibfnamefont {N.}~\bibnamefont
			{Pandey}}, \bibinfo {author} {\bibfnamefont {M.}~\bibnamefont {Gupta}},
		\bibinfo {author} {\bibfnamefont {R.}~\bibnamefont {Gupta}}, \bibinfo
		{author} {\bibfnamefont {S.}~\bibnamefont {Amir}}, \ and\ \bibinfo {author}
		{\bibfnamefont {J.}~\bibnamefont {Stahn}},\ }\href@noop {} {\bibfield
		{journal} {\bibinfo  {journal} {Applied Physics A}\ }\textbf {\bibinfo
			{volume} {125}},\ \bibinfo {pages} {539} (\bibinfo {year}
		{2019}{\natexlab{a}})}\BibitemShut {NoStop}%
	\bibitem [{\citenamefont {Pandey}\ \emph {et~al.}(2018)\citenamefont {Pandey},
		\citenamefont {Gupta}, \citenamefont {Gupta}, \citenamefont {Rajput},\ and\
		\citenamefont {Stahn}}]{JMMM18_Co4N_NP}%
	\BibitemOpen
	\bibfield  {author} {\bibinfo {author} {\bibfnamefont {N.}~\bibnamefont
			{Pandey}}, \bibinfo {author} {\bibfnamefont {M.}~\bibnamefont {Gupta}},
		\bibinfo {author} {\bibfnamefont {R.}~\bibnamefont {Gupta}}, \bibinfo
		{author} {\bibfnamefont {P.}~\bibnamefont {Rajput}}, \ and\ \bibinfo {author}
		{\bibfnamefont {J.}~\bibnamefont {Stahn}},\ }\href@noop {} {\bibfield
		{journal} {\bibinfo  {journal} {Journal of Magnetism and Magnetic Materials}\
		}\textbf {\bibinfo {volume} {448}},\ \bibinfo {pages} {274} (\bibinfo {year}
		{2018})}\BibitemShut {NoStop}%
	\bibitem [{\citenamefont {Pandey}\ \emph {et~al.}(2017)\citenamefont {Pandey},
		\citenamefont {Gupta}, \citenamefont {Gupta}, \citenamefont {Chakravarty},
		\citenamefont {Shukla},\ and\ \citenamefont {Devishvili}}]{JAC16_NPandey}%
	\BibitemOpen
	\bibfield  {author} {\bibinfo {author} {\bibfnamefont {N.}~\bibnamefont
			{Pandey}}, \bibinfo {author} {\bibfnamefont {M.}~\bibnamefont {Gupta}},
		\bibinfo {author} {\bibfnamefont {R.}~\bibnamefont {Gupta}}, \bibinfo
		{author} {\bibfnamefont {S.}~\bibnamefont {Chakravarty}}, \bibinfo {author}
		{\bibfnamefont {N.}~\bibnamefont {Shukla}}, \ and\ \bibinfo {author}
		{\bibfnamefont {A.}~\bibnamefont {Devishvili}},\ }\href {\doibase
		https://doi.org/10.1016/j.jallcom.2016.10.095} {\bibfield  {journal}
		{\bibinfo  {journal} {Journal of Alloys and Compounds}\ }\textbf {\bibinfo
			{volume} {694}},\ \bibinfo {pages} {1209 } (\bibinfo {year}
		{2017})}\BibitemShut {NoStop}%
	\bibitem [{\citenamefont {Pandey}\ \emph
		{et~al.}(2019{\natexlab{b}})\citenamefont {Pandey}, \citenamefont {Gupta},
		\citenamefont {Gupta}, \citenamefont {Hussain}, \citenamefont {Reddy},
		\citenamefont {Phase},\ and\ \citenamefont {Stahn}}]{PRB19_Co4N_NP}%
	\BibitemOpen
	\bibfield  {author} {\bibinfo {author} {\bibfnamefont {N.}~\bibnamefont
			{Pandey}}, \bibinfo {author} {\bibfnamefont {M.}~\bibnamefont {Gupta}},
		\bibinfo {author} {\bibfnamefont {R.}~\bibnamefont {Gupta}}, \bibinfo
		{author} {\bibfnamefont {Z.}~\bibnamefont {Hussain}}, \bibinfo {author}
		{\bibfnamefont {V.}~\bibnamefont {Reddy}}, \bibinfo {author} {\bibfnamefont
			{D.}~\bibnamefont {Phase}}, \ and\ \bibinfo {author} {\bibfnamefont
			{J.}~\bibnamefont {Stahn}},\ }\href@noop {} {\bibfield  {journal} {\bibinfo
			{journal} {Physical Review B}\ }\textbf {\bibinfo {volume} {99}},\ \bibinfo
		{pages} {214109} (\bibinfo {year} {2019}{\natexlab{b}})}\BibitemShut
	{NoStop}%
	\bibitem [{\citenamefont {Asahara}\ \emph {et~al.}(2001)\citenamefont
		{Asahara}, \citenamefont {Migita}, \citenamefont {Tanaka},\ and\
		\citenamefont {Kawabata}}]{Vac01_Asahara}%
	\BibitemOpen
	\bibfield  {author} {\bibinfo {author} {\bibfnamefont {H.}~\bibnamefont
			{Asahara}}, \bibinfo {author} {\bibfnamefont {T.}~\bibnamefont {Migita}},
		\bibinfo {author} {\bibfnamefont {T.}~\bibnamefont {Tanaka}}, \ and\ \bibinfo
		{author} {\bibfnamefont {K.}~\bibnamefont {Kawabata}},\ }\href {\doibase
		http://dx.doi.org/10.1016/S0042-207X(00)00453-X} {\bibfield  {journal}
		{\bibinfo  {journal} {Vacuum}\ }\textbf {\bibinfo {volume} {62}},\ \bibinfo
		{pages} {293 } (\bibinfo {year} {2001})}\BibitemShut {NoStop}%
	\bibitem [{\citenamefont {Fang}\ \emph {et~al.}(2004)\citenamefont {Fang},
		\citenamefont {Yang}, \citenamefont {Hsu}, \citenamefont {Chen},
		\citenamefont {Lin},\ and\ \citenamefont {Chen}}]{JVSTA04_Fang}%
	\BibitemOpen
	\bibfield  {author} {\bibinfo {author} {\bibfnamefont {J.-S.}\ \bibnamefont
			{Fang}}, \bibinfo {author} {\bibfnamefont {L.-C.}\ \bibnamefont {Yang}},
		\bibinfo {author} {\bibfnamefont {C.-S.}\ \bibnamefont {Hsu}}, \bibinfo
		{author} {\bibfnamefont {G.-S.}\ \bibnamefont {Chen}}, \bibinfo {author}
		{\bibfnamefont {Y.-W.}\ \bibnamefont {Lin}}, \ and\ \bibinfo {author}
		{\bibfnamefont {G.-S.}\ \bibnamefont {Chen}},\ }\href@noop {} {\bibfield
		{journal} {\bibinfo  {journal} {Journal of Vacuum Science \& Technology A:
				Vacuum, Surfaces, and Films}\ }\textbf {\bibinfo {volume} {22}},\ \bibinfo
		{pages} {698} (\bibinfo {year} {2004})}\BibitemShut {NoStop}%
	\bibitem [{\citenamefont {Zhou}\ \emph {et~al.}(2017)\citenamefont {Zhou},
		\citenamefont {Cao}, \citenamefont {Wang}, \citenamefont {Hao}, \citenamefont
		{Hou}, \citenamefont {Qu}, \citenamefont {Du}, \citenamefont {Asiri},
		\citenamefont {Zheng},\ and\ \citenamefont {Sun}}]{EupChem17_Co2N_Zhou}%
	\BibitemOpen
	\bibfield  {author} {\bibinfo {author} {\bibfnamefont {D.}~\bibnamefont
			{Zhou}}, \bibinfo {author} {\bibfnamefont {X.}~\bibnamefont {Cao}}, \bibinfo
		{author} {\bibfnamefont {Z.}~\bibnamefont {Wang}}, \bibinfo {author}
		{\bibfnamefont {S.}~\bibnamefont {Hao}}, \bibinfo {author} {\bibfnamefont
			{X.}~\bibnamefont {Hou}}, \bibinfo {author} {\bibfnamefont {F.}~\bibnamefont
			{Qu}}, \bibinfo {author} {\bibfnamefont {G.}~\bibnamefont {Du}}, \bibinfo
		{author} {\bibfnamefont {A.~M.}\ \bibnamefont {Asiri}}, \bibinfo {author}
		{\bibfnamefont {C.}~\bibnamefont {Zheng}}, \ and\ \bibinfo {author}
		{\bibfnamefont {X.}~\bibnamefont {Sun}},\ }\href@noop {} {\bibfield
		{journal} {\bibinfo  {journal} {Chemistry--A European Journal}\ }\textbf
		{\bibinfo {volume} {23}},\ \bibinfo {pages} {5214} (\bibinfo {year}
		{2017})}\BibitemShut {NoStop}%
	\bibitem [{\citenamefont {Guo}\ \emph {et~al.}(2019)\citenamefont {Guo},
		\citenamefont {Tian}, \citenamefont {Wang}, \citenamefont {Ke},\ and\
		\citenamefont {Zhu}}]{ASS19_Co2N_Guo_ORR}%
	\BibitemOpen
	\bibfield  {author} {\bibinfo {author} {\bibfnamefont {D.}~\bibnamefont
			{Guo}}, \bibinfo {author} {\bibfnamefont {Z.}~\bibnamefont {Tian}}, \bibinfo
		{author} {\bibfnamefont {J.}~\bibnamefont {Wang}}, \bibinfo {author}
		{\bibfnamefont {X.}~\bibnamefont {Ke}}, \ and\ \bibinfo {author}
		{\bibfnamefont {Y.}~\bibnamefont {Zhu}},\ }\href@noop {} {\bibfield
		{journal} {\bibinfo  {journal} {Applied Surface Science}\ }\textbf {\bibinfo
			{volume} {473}},\ \bibinfo {pages} {555} (\bibinfo {year}
		{2019})}\BibitemShut {NoStop}%
	\bibitem [{\citenamefont {Suzuki}\ \emph {et~al.}(1995)\citenamefont {Suzuki},
		\citenamefont {Kaneko}, \citenamefont {Yoshida}, \citenamefont {Morita},\
		and\ \citenamefont {Fujimori}}]{JAC95_Suzuki}%
	\BibitemOpen
	\bibfield  {author} {\bibinfo {author} {\bibfnamefont {K.}~\bibnamefont
			{Suzuki}}, \bibinfo {author} {\bibfnamefont {T.}~\bibnamefont {Kaneko}},
		\bibinfo {author} {\bibfnamefont {H.}~\bibnamefont {Yoshida}}, \bibinfo
		{author} {\bibfnamefont {H.}~\bibnamefont {Morita}}, \ and\ \bibinfo {author}
		{\bibfnamefont {H.}~\bibnamefont {Fujimori}},\ }\href {\doibase
		http://dx.doi.org/10.1016/0925-8388(95)01561-2} {\bibfield  {journal}
		{\bibinfo  {journal} {Journal of Alloys and Compounds}\ }\textbf {\bibinfo
			{volume} {224}},\ \bibinfo {pages} {232 } (\bibinfo {year}
		{1995})}\BibitemShut {NoStop}%
	\bibitem [{\citenamefont {Shi}\ \emph {et~al.}(2008)\citenamefont {Shi},
		\citenamefont {Wan}, \citenamefont {Zhang},\ and\ \citenamefont
		{Zhao}}]{AFM08_CoN}%
	\BibitemOpen
	\bibfield  {author} {\bibinfo {author} {\bibfnamefont {Y.}~\bibnamefont
			{Shi}}, \bibinfo {author} {\bibfnamefont {Y.}~\bibnamefont {Wan}}, \bibinfo
		{author} {\bibfnamefont {R.}~\bibnamefont {Zhang}}, \ and\ \bibinfo {author}
		{\bibfnamefont {D.}~\bibnamefont {Zhao}},\ }\href@noop {} {\bibfield
		{journal} {\bibinfo  {journal} {Advanced Functional Materials}\ }\textbf
		{\bibinfo {volume} {18}},\ \bibinfo {pages} {2436} (\bibinfo {year}
		{2008})}\BibitemShut {NoStop}%
	\bibitem [{\citenamefont {Xie}\ \emph {et~al.}(2018)\citenamefont {Xie},
		\citenamefont {Liu}, \citenamefont {Xie}, \citenamefont {Sun},\ and\
		\citenamefont {Luo}}]{Ni-N_nanosheet_metallic_18}%
	\BibitemOpen
	\bibfield  {author} {\bibinfo {author} {\bibfnamefont {F.}~\bibnamefont
			{Xie}}, \bibinfo {author} {\bibfnamefont {T.}~\bibnamefont {Liu}}, \bibinfo
		{author} {\bibfnamefont {L.}~\bibnamefont {Xie}}, \bibinfo {author}
		{\bibfnamefont {X.}~\bibnamefont {Sun}}, \ and\ \bibinfo {author}
		{\bibfnamefont {Y.}~\bibnamefont {Luo}},\ }\href@noop {} {\bibfield
		{journal} {\bibinfo  {journal} {Sensors and Actuators B: Chemical}\ }\textbf
		{\bibinfo {volume} {255}},\ \bibinfo {pages} {2794} (\bibinfo {year}
		{2018})}\BibitemShut {NoStop}%
	\bibitem [{\citenamefont {Yu}\ \emph {et~al.}(2015)\citenamefont {Yu},
		\citenamefont {Gao}, \citenamefont {Shen}, \citenamefont {Zheng},
		\citenamefont {Wu}, \citenamefont {Wang}, \citenamefont {Song},\ and\
		\citenamefont {Ding}}]{Ni3N_Electrode_JMCA15}%
	\BibitemOpen
	\bibfield  {author} {\bibinfo {author} {\bibfnamefont {Y.}~\bibnamefont
			{Yu}}, \bibinfo {author} {\bibfnamefont {W.}~\bibnamefont {Gao}}, \bibinfo
		{author} {\bibfnamefont {Z.}~\bibnamefont {Shen}}, \bibinfo {author}
		{\bibfnamefont {Q.}~\bibnamefont {Zheng}}, \bibinfo {author} {\bibfnamefont
			{H.}~\bibnamefont {Wu}}, \bibinfo {author} {\bibfnamefont {X.}~\bibnamefont
			{Wang}}, \bibinfo {author} {\bibfnamefont {W.}~\bibnamefont {Song}}, \ and\
		\bibinfo {author} {\bibfnamefont {K.}~\bibnamefont {Ding}},\ }\href@noop {}
	{\bibfield  {journal} {\bibinfo  {journal} {Journal of Materials Chemistry
				A}\ }\textbf {\bibinfo {volume} {3}},\ \bibinfo {pages} {16633} (\bibinfo
		{year} {2015})}\BibitemShut {NoStop}%
	\bibitem [{\citenamefont {Li}\ \emph {et~al.}(2019)\citenamefont {Li},
		\citenamefont {Pu}, \citenamefont {Taiki}, \citenamefont {Zhang},
		\citenamefont {Xie}, \citenamefont {Fujiwara}, \citenamefont {Kitahata},
		\citenamefont {Li}, \citenamefont {Kobayashi}, \citenamefont {Ito} \emph
		{et~al.}}]{Ni-N_GaN_anode_Li_Vac19}%
	\BibitemOpen
	\bibfield  {author} {\bibinfo {author} {\bibfnamefont {X.}~\bibnamefont
			{Li}}, \bibinfo {author} {\bibfnamefont {T.}~\bibnamefont {Pu}}, \bibinfo
		{author} {\bibfnamefont {H.}~\bibnamefont {Taiki}}, \bibinfo {author}
		{\bibfnamefont {T.}~\bibnamefont {Zhang}}, \bibinfo {author} {\bibfnamefont
			{T.}~\bibnamefont {Xie}}, \bibinfo {author} {\bibfnamefont {S.~J.~L.}\
			\bibnamefont {Fujiwara}}, \bibinfo {author} {\bibfnamefont {H.}~\bibnamefont
			{Kitahata}}, \bibinfo {author} {\bibfnamefont {L.}~\bibnamefont {Li}},
		\bibinfo {author} {\bibfnamefont {S.}~\bibnamefont {Kobayashi}}, \bibinfo
		{author} {\bibfnamefont {M.}~\bibnamefont {Ito}},  \emph {et~al.},\
	}\href@noop {} {\bibfield  {journal} {\bibinfo  {journal} {Vacuum}\ }\textbf
		{\bibinfo {volume} {162}},\ \bibinfo {pages} {72} (\bibinfo {year}
		{2019})}\BibitemShut {NoStop}%
	\bibitem [{\citenamefont {Gillot}\ \emph {et~al.}(2011)\citenamefont {Gillot},
		\citenamefont {Oró-Solé},\ and\ \citenamefont
		{Palacín}}]{Ni-N_lithium_electrodes_Gillot11}%
	\BibitemOpen
	\bibfield  {author} {\bibinfo {author} {\bibfnamefont {F.}~\bibnamefont
			{Gillot}}, \bibinfo {author} {\bibfnamefont {J.}~\bibnamefont {Oró-Solé}},
		\ and\ \bibinfo {author} {\bibfnamefont {M.~R.}\ \bibnamefont {Palacín}},\
	}\href {\doibase 10.1039/C0JM04144K} {\bibfield  {journal} {\bibinfo
			{journal} {J. Mater. Chem.}\ }\textbf {\bibinfo {volume} {21}},\ \bibinfo
		{pages} {9997} (\bibinfo {year} {2011})}\BibitemShut {NoStop}%
	\bibitem [{\citenamefont {Balogun}\ \emph {et~al.}(2015)\citenamefont
		{Balogun}, \citenamefont {Qiu}, \citenamefont {Wang}, \citenamefont {Fang},
		\citenamefont {Lu},\ and\ \citenamefont
		{Tong}}]{TMN_electrode_Balogun_RSC15}%
	\BibitemOpen
	\bibfield  {author} {\bibinfo {author} {\bibfnamefont {M.-S.}\ \bibnamefont
			{Balogun}}, \bibinfo {author} {\bibfnamefont {W.}~\bibnamefont {Qiu}},
		\bibinfo {author} {\bibfnamefont {W.}~\bibnamefont {Wang}}, \bibinfo {author}
		{\bibfnamefont {P.}~\bibnamefont {Fang}}, \bibinfo {author} {\bibfnamefont
			{X.}~\bibnamefont {Lu}}, \ and\ \bibinfo {author} {\bibfnamefont
			{Y.}~\bibnamefont {Tong}},\ }\href@noop {} {\bibfield  {journal} {\bibinfo
			{journal} {Journal of materials chemistry A}\ }\textbf {\bibinfo {volume}
			{3}},\ \bibinfo {pages} {1364} (\bibinfo {year} {2015})}\BibitemShut
	{NoStop}%
	\bibitem [{\citenamefont {Jia}\ \emph {et~al.}(2019)\citenamefont {Jia},
		\citenamefont {Ma}, \citenamefont {Li}, \citenamefont {He}, \citenamefont
		{Shao},\ and\ \citenamefont {Zhang}}]{FMS19_NiO_Ni2N_Lit_ion_Bat}%
	\BibitemOpen
	\bibfield  {author} {\bibinfo {author} {\bibfnamefont {Y.}~\bibnamefont
			{Jia}}, \bibinfo {author} {\bibfnamefont {Z.}~\bibnamefont {Ma}}, \bibinfo
		{author} {\bibfnamefont {Z.}~\bibnamefont {Li}}, \bibinfo {author}
		{\bibfnamefont {Z.}~\bibnamefont {He}}, \bibinfo {author} {\bibfnamefont
			{J.}~\bibnamefont {Shao}}, \ and\ \bibinfo {author} {\bibfnamefont
			{H.}~\bibnamefont {Zhang}},\ }\href@noop {} {\bibfield  {journal} {\bibinfo
			{journal} {Frontiers of Materials Science}\ ,\ \bibinfo {pages} {1}}
		(\bibinfo {year} {2019})}\BibitemShut {NoStop}%
	\bibitem [{\citenamefont {Kang}\ \emph {et~al.}(2015)\citenamefont {Kang},
		\citenamefont {Park}, \citenamefont {Kim}, \citenamefont {Park},
		\citenamefont {Chung}, \citenamefont {Yu}, \citenamefont {Kim}, \citenamefont
		{Park}, \citenamefont {Choi}, \citenamefont {Lee} \emph
		{et~al.}}]{Ni-N_dye_electrode_Kang_SciRep15}%
	\BibitemOpen
	\bibfield  {author} {\bibinfo {author} {\bibfnamefont {J.~S.}\ \bibnamefont
			{Kang}}, \bibinfo {author} {\bibfnamefont {M.-A.}\ \bibnamefont {Park}},
		\bibinfo {author} {\bibfnamefont {J.-Y.}\ \bibnamefont {Kim}}, \bibinfo
		{author} {\bibfnamefont {S.~H.}\ \bibnamefont {Park}}, \bibinfo {author}
		{\bibfnamefont {D.~Y.}\ \bibnamefont {Chung}}, \bibinfo {author}
		{\bibfnamefont {S.-H.}\ \bibnamefont {Yu}}, \bibinfo {author} {\bibfnamefont
			{J.}~\bibnamefont {Kim}}, \bibinfo {author} {\bibfnamefont {J.}~\bibnamefont
			{Park}}, \bibinfo {author} {\bibfnamefont {J.-W.}\ \bibnamefont {Choi}},
		\bibinfo {author} {\bibfnamefont {K.~J.}\ \bibnamefont {Lee}},  \emph
		{et~al.},\ }\href@noop {} {\bibfield  {journal} {\bibinfo  {journal}
			{Scientific reports}\ }\textbf {\bibinfo {volume} {5}},\ \bibinfo {pages}
		{10450} (\bibinfo {year} {2015})}\BibitemShut {NoStop}%
	\bibitem [{\citenamefont {Keraudy}\ \emph {et~al.}(2019)\citenamefont
		{Keraudy}, \citenamefont {Athouel}, \citenamefont {Hamon}, \citenamefont
		{Girault}, \citenamefont {Gloaguen}, \citenamefont {Richard-Plouet},\ and\
		\citenamefont {Jouan}}]{Ni-N_dcMS_Keraudy_TSF19}%
	\BibitemOpen
	\bibfield  {author} {\bibinfo {author} {\bibfnamefont {J.}~\bibnamefont
			{Keraudy}}, \bibinfo {author} {\bibfnamefont {L.}~\bibnamefont {Athouel}},
		\bibinfo {author} {\bibfnamefont {J.}~\bibnamefont {Hamon}}, \bibinfo
		{author} {\bibfnamefont {B.}~\bibnamefont {Girault}}, \bibinfo {author}
		{\bibfnamefont {D.}~\bibnamefont {Gloaguen}}, \bibinfo {author}
		{\bibfnamefont {M.}~\bibnamefont {Richard-Plouet}}, \ and\ \bibinfo {author}
		{\bibfnamefont {P.-Y.}\ \bibnamefont {Jouan}},\ }\href@noop {} {\bibfield
		{journal} {\bibinfo  {journal} {Thin Solid Films}\ }\textbf {\bibinfo
			{volume} {669}},\ \bibinfo {pages} {659} (\bibinfo {year}
		{2019})}\BibitemShut {NoStop}%
	\bibitem [{\citenamefont {Gage}\ \emph {et~al.}(2016)\citenamefont {Gage},
		\citenamefont {Trewyn}, \citenamefont {Ciobanu}, \citenamefont {Pylypenko},\
		and\ \citenamefont {Richards}}]{Ni3N_Catalytic_Gage16}%
	\BibitemOpen
	\bibfield  {author} {\bibinfo {author} {\bibfnamefont {S.~H.}\ \bibnamefont
			{Gage}}, \bibinfo {author} {\bibfnamefont {B.}~\bibnamefont {Trewyn}},
		\bibinfo {author} {\bibfnamefont {C.}~\bibnamefont {Ciobanu}}, \bibinfo
		{author} {\bibfnamefont {S.}~\bibnamefont {Pylypenko}}, \ and\ \bibinfo
		{author} {\bibfnamefont {R.}~\bibnamefont {Richards}},\ }\href@noop {}
	{\bibfield  {journal} {\bibinfo  {journal} {Catalysis Science \& Technology}\
		}\textbf {\bibinfo {volume} {6}},\ \bibinfo {pages} {4059} (\bibinfo {year}
		{2016})}\BibitemShut {NoStop}%
	\bibitem [{\citenamefont {Kim}\ \emph {et~al.}(2012)\citenamefont {Kim},
		\citenamefont {An},\ and\ \citenamefont {Kim}}]{Ni-N_metal_trans_Kim_IEEE12}%
	\BibitemOpen
	\bibfield  {author} {\bibinfo {author} {\bibfnamefont {H.-D.}\ \bibnamefont
			{Kim}}, \bibinfo {author} {\bibfnamefont {H.-M.}\ \bibnamefont {An}}, \ and\
		\bibinfo {author} {\bibfnamefont {T.~G.}\ \bibnamefont {Kim}},\ }\href@noop
	{} {\bibfield  {journal} {\bibinfo  {journal} {IEEE Transactions on Electron
				Devices}\ }\textbf {\bibinfo {volume} {59}},\ \bibinfo {pages} {2302}
		(\bibinfo {year} {2012})}\BibitemShut {NoStop}%
	\bibitem [{\citenamefont {Yun}\ \emph {et~al.}(2014)\citenamefont {Yun},
		\citenamefont {Kim}, \citenamefont {Man~Hong}, \citenamefont {Hyun~Park},
		\citenamefont {Su~Jeon},\ and\ \citenamefont
		{Geun~Kim}}]{Ni-N_metal_trans_JAP14}%
	\BibitemOpen
	\bibfield  {author} {\bibinfo {author} {\bibfnamefont {M.~J.}\ \bibnamefont
			{Yun}}, \bibinfo {author} {\bibfnamefont {H.-D.}\ \bibnamefont {Kim}},
		\bibinfo {author} {\bibfnamefont {S.}~\bibnamefont {Man~Hong}}, \bibinfo
		{author} {\bibfnamefont {J.}~\bibnamefont {Hyun~Park}}, \bibinfo {author}
		{\bibfnamefont {D.}~\bibnamefont {Su~Jeon}}, \ and\ \bibinfo {author}
		{\bibfnamefont {T.}~\bibnamefont {Geun~Kim}},\ }\href@noop {} {\bibfield
		{journal} {\bibinfo  {journal} {Journal of Applied Physics}\ }\textbf
		{\bibinfo {volume} {115}},\ \bibinfo {pages} {094305} (\bibinfo {year}
		{2014})}\BibitemShut {NoStop}%
	\bibitem [{\citenamefont {Ju~Yun}\ \emph {et~al.}(2013)\citenamefont {Ju~Yun},
		\citenamefont {Kim},\ and\ \citenamefont
		{Geun~Kim}}]{Ni-N_film_metal_trans_JVST13}%
	\BibitemOpen
	\bibfield  {author} {\bibinfo {author} {\bibfnamefont {M.}~\bibnamefont
			{Ju~Yun}}, \bibinfo {author} {\bibfnamefont {H.-D.}\ \bibnamefont {Kim}}, \
		and\ \bibinfo {author} {\bibfnamefont {T.}~\bibnamefont {Geun~Kim}},\
	}\href@noop {} {\bibfield  {journal} {\bibinfo  {journal} {Journal of Vacuum
				Science \& Technology B, Nanotechnology and Microelectronics: Materials,
				Processing, Measurement, and Phenomena}\ }\textbf {\bibinfo {volume} {31}},\
		\bibinfo {pages} {060601} (\bibinfo {year} {2013})}\BibitemShut {NoStop}%
	\bibitem [{\citenamefont {Kim}\ \emph {et~al.}(2014)\citenamefont {Kim},
		\citenamefont {Yun}, \citenamefont {Hong},\ and\ \citenamefont
		{Kim}}]{Ni-N_metal_trans_ASP14}%
	\BibitemOpen
	\bibfield  {author} {\bibinfo {author} {\bibfnamefont {H.-D.}\ \bibnamefont
			{Kim}}, \bibinfo {author} {\bibfnamefont {M.~J.}\ \bibnamefont {Yun}},
		\bibinfo {author} {\bibfnamefont {S.~M.}\ \bibnamefont {Hong}}, \ and\
		\bibinfo {author} {\bibfnamefont {T.~G.}\ \bibnamefont {Kim}},\ }\href@noop
	{} {\bibfield  {journal} {\bibinfo  {journal} {Journal of nanoscience and
				nanotechnology}\ }\textbf {\bibinfo {volume} {14}},\ \bibinfo {pages} {9088}
		(\bibinfo {year} {2014})}\BibitemShut {NoStop}%
	\bibitem [{\citenamefont {Guillermet}\ and\ \citenamefont
		{Frisk}(1991)}]{Ni-N_thermal_stab_phase_dia_Guillermet1991}%
	\BibitemOpen
	\bibfield  {author} {\bibinfo {author} {\bibfnamefont {A.~F.}\ \bibnamefont
			{Guillermet}}\ and\ \bibinfo {author} {\bibfnamefont {K.}~\bibnamefont
			{Frisk}},\ }\href@noop {} {\bibfield  {journal} {\bibinfo  {journal}
			{International journal of thermophysics}\ }\textbf {\bibinfo {volume} {12}},\
		\bibinfo {pages} {417} (\bibinfo {year} {1991})}\BibitemShut {NoStop}%
	\bibitem [{\citenamefont {Neklyudov}\ and\ \citenamefont
		{Morozov}(2004)}]{Phase_diagram_N_imp_Ni_Neklyudov04}%
	\BibitemOpen
	\bibfield  {author} {\bibinfo {author} {\bibfnamefont {I.}~\bibnamefont
			{Neklyudov}}\ and\ \bibinfo {author} {\bibfnamefont {A.}~\bibnamefont
			{Morozov}},\ }\href@noop {} {\bibfield  {journal} {\bibinfo  {journal}
			{Physica B: Condensed Matter}\ }\textbf {\bibinfo {volume} {350}},\ \bibinfo
		{pages} {325} (\bibinfo {year} {2004})}\BibitemShut {NoStop}%
	\bibitem [{\citenamefont {Terao}(1959)}]{Ni4N_2_Terao59}%
	\BibitemOpen
	\bibfield  {author} {\bibinfo {author} {\bibfnamefont {N.}~\bibnamefont
			{Terao}},\ }\href@noop {} {\bibfield  {journal} {\bibinfo  {journal}
			{Naturwissenschaften}\ }\textbf {\bibinfo {volume} {46}},\ \bibinfo {pages}
		{204} (\bibinfo {year} {1959})}\BibitemShut {NoStop}%
	\bibitem [{\citenamefont {Dorman}\ and\ \citenamefont
		{Sikkens}(1983)}]{Ni4N_2_Dorman_TSF83}%
	\BibitemOpen
	\bibfield  {author} {\bibinfo {author} {\bibfnamefont {G.}~\bibnamefont
			{Dorman}}\ and\ \bibinfo {author} {\bibfnamefont {M.}~\bibnamefont
			{Sikkens}},\ }\href@noop {} {\bibfield  {journal} {\bibinfo  {journal} {Thin
				Solid Films}\ }\textbf {\bibinfo {volume} {105}},\ \bibinfo {pages} {251}
		(\bibinfo {year} {1983})}\BibitemShut {NoStop}%
	\bibitem [{\citenamefont {Hemzalov{\'a}}\ \emph {et~al.}(2013)\citenamefont
		{Hemzalov{\'a}}, \citenamefont {Fri{\'a}k}, \citenamefont {{\v{S}}ob},
		\citenamefont {Ma}, \citenamefont {Udyansky}, \citenamefont {Raabe},\ and\
		\citenamefont {Neugebauer}}]{Ni4N_allotropes_Hemzalova_PRB13}%
	\BibitemOpen
	\bibfield  {author} {\bibinfo {author} {\bibfnamefont {P.}~\bibnamefont
			{Hemzalov{\'a}}}, \bibinfo {author} {\bibfnamefont {M.}~\bibnamefont
			{Fri{\'a}k}}, \bibinfo {author} {\bibfnamefont {M.}~\bibnamefont
			{{\v{S}}ob}}, \bibinfo {author} {\bibfnamefont {D.}~\bibnamefont {Ma}},
		\bibinfo {author} {\bibfnamefont {A.}~\bibnamefont {Udyansky}}, \bibinfo
		{author} {\bibfnamefont {D.}~\bibnamefont {Raabe}}, \ and\ \bibinfo {author}
		{\bibfnamefont {J.}~\bibnamefont {Neugebauer}},\ }\href@noop {} {\bibfield
		{journal} {\bibinfo  {journal} {Physical Review B}\ }\textbf {\bibinfo
			{volume} {88}},\ \bibinfo {pages} {174103} (\bibinfo {year}
		{2013})}\BibitemShut {NoStop}%
	\bibitem [{\citenamefont {Fang}\ \emph {et~al.}(2014)\citenamefont {Fang},
		\citenamefont {Koster}, \citenamefont {Li},\ and\ \citenamefont {van
			Huis}}]{TM4N_2_Fang_RSC14}%
	\BibitemOpen
	\bibfield  {author} {\bibinfo {author} {\bibfnamefont {C.-M.}\ \bibnamefont
			{Fang}}, \bibinfo {author} {\bibfnamefont {R.~S.}\ \bibnamefont {Koster}},
		\bibinfo {author} {\bibfnamefont {W.-F.}\ \bibnamefont {Li}}, \ and\ \bibinfo
		{author} {\bibfnamefont {M.~A.}\ \bibnamefont {van Huis}},\ }\href@noop {}
	{\bibfield  {journal} {\bibinfo  {journal} {RSC Advances}\ }\textbf {\bibinfo
			{volume} {4}},\ \bibinfo {pages} {7885} (\bibinfo {year} {2014})}\BibitemShut
	{NoStop}%
	\bibitem [{\citenamefont {Leineweber}\ and\ \citenamefont
		{Maisel}(2019)}]{Ni4N_Leineweber19}%
	\BibitemOpen
	\bibfield  {author} {\bibinfo {author} {\bibfnamefont {A.}~\bibnamefont
			{Leineweber}}\ and\ \bibinfo {author} {\bibfnamefont {S.}~\bibnamefont
			{Maisel}},\ }\href@noop {} {\bibfield  {journal} {\bibinfo  {journal}
			{Computational Materials Science}\ }\textbf {\bibinfo {volume} {161}},\
		\bibinfo {pages} {209} (\bibinfo {year} {2019})}\BibitemShut {NoStop}%
	\bibitem [{\citenamefont {Vempaire}\ \emph
		{et~al.}(2009{\natexlab{a}})\citenamefont {Vempaire}, \citenamefont
		{Miraglia}, \citenamefont {Pelletier}, \citenamefont {Fruchart},
		\citenamefont {Hlil}, \citenamefont {Ortega}, \citenamefont {Sulpice},\ and\
		\citenamefont {Fettar}}]{JAC09_Ni3N_film_Vempaire}%
	\BibitemOpen
	\bibfield  {author} {\bibinfo {author} {\bibfnamefont {D.}~\bibnamefont
			{Vempaire}}, \bibinfo {author} {\bibfnamefont {S.}~\bibnamefont {Miraglia}},
		\bibinfo {author} {\bibfnamefont {J.}~\bibnamefont {Pelletier}}, \bibinfo
		{author} {\bibfnamefont {D.}~\bibnamefont {Fruchart}}, \bibinfo {author}
		{\bibfnamefont {E.}~\bibnamefont {Hlil}}, \bibinfo {author} {\bibfnamefont
			{L.}~\bibnamefont {Ortega}}, \bibinfo {author} {\bibfnamefont
			{A.}~\bibnamefont {Sulpice}}, \ and\ \bibinfo {author} {\bibfnamefont
			{F.}~\bibnamefont {Fettar}},\ }\href {\doibase
		https://doi.org/10.1016/j.jallcom.2009.02.066} {\bibfield  {journal}
		{\bibinfo  {journal} {Journal of Alloys and Compounds}\ }\textbf {\bibinfo
			{volume} {480}},\ \bibinfo {pages} {225 } (\bibinfo {year}
		{2009}{\natexlab{a}})}\BibitemShut {NoStop}%
	\bibitem [{\citenamefont {Vempaire}\ \emph
		{et~al.}(2009{\natexlab{b}})\citenamefont {Vempaire}, \citenamefont {Fettar},
		\citenamefont {Ortega}, \citenamefont {Pierre}, \citenamefont {Miraglia},
		\citenamefont {Sulpice}, \citenamefont {Pelletier}, \citenamefont {Hlil},\
		and\ \citenamefont {Fruchart}}]{JAP09_Ni3N_film_Vempaire}%
	\BibitemOpen
	\bibfield  {author} {\bibinfo {author} {\bibfnamefont {D.}~\bibnamefont
			{Vempaire}}, \bibinfo {author} {\bibfnamefont {F.}~\bibnamefont {Fettar}},
		\bibinfo {author} {\bibfnamefont {L.}~\bibnamefont {Ortega}}, \bibinfo
		{author} {\bibfnamefont {F.}~\bibnamefont {Pierre}}, \bibinfo {author}
		{\bibfnamefont {S.}~\bibnamefont {Miraglia}}, \bibinfo {author}
		{\bibfnamefont {A.}~\bibnamefont {Sulpice}}, \bibinfo {author} {\bibfnamefont
			{J.}~\bibnamefont {Pelletier}}, \bibinfo {author} {\bibfnamefont {E.-K.}\
			\bibnamefont {Hlil}}, \ and\ \bibinfo {author} {\bibfnamefont
			{D.}~\bibnamefont {Fruchart}},\ }\href@noop {} {\bibfield  {journal}
		{\bibinfo  {journal} {Journal of Applied Physics}\ }\textbf {\bibinfo
			{volume} {106}},\ \bibinfo {pages} {073911} (\bibinfo {year}
		{2009}{\natexlab{b}})}\BibitemShut {NoStop}%
	\bibitem [{\citenamefont {Nishihara}\ \emph {et~al.}(2014)\citenamefont
		{Nishihara}, \citenamefont {Suzuki}, \citenamefont {Umetsu}, \citenamefont
		{Kanomata}, \citenamefont {Kaneko}, \citenamefont {Zhou}, \citenamefont
		{Tsujikawa}, \citenamefont {Shirai}, \citenamefont {Sakon}, \citenamefont
		{Wada} \emph {et~al.}}]{PhyB14_Ni2N_film_Nishihara}%
	\BibitemOpen
	\bibfield  {author} {\bibinfo {author} {\bibfnamefont {H.}~\bibnamefont
			{Nishihara}}, \bibinfo {author} {\bibfnamefont {K.}~\bibnamefont {Suzuki}},
		\bibinfo {author} {\bibfnamefont {R.}~\bibnamefont {Umetsu}}, \bibinfo
		{author} {\bibfnamefont {T.}~\bibnamefont {Kanomata}}, \bibinfo {author}
		{\bibfnamefont {T.}~\bibnamefont {Kaneko}}, \bibinfo {author} {\bibfnamefont
			{M.}~\bibnamefont {Zhou}}, \bibinfo {author} {\bibfnamefont {M.}~\bibnamefont
			{Tsujikawa}}, \bibinfo {author} {\bibfnamefont {M.}~\bibnamefont {Shirai}},
		\bibinfo {author} {\bibfnamefont {T.}~\bibnamefont {Sakon}}, \bibinfo
		{author} {\bibfnamefont {T.}~\bibnamefont {Wada}},  \emph {et~al.},\
	}\href@noop {} {\bibfield  {journal} {\bibinfo  {journal} {Physica B:
				Condensed Matter}\ }\textbf {\bibinfo {volume} {449}},\ \bibinfo {pages} {85}
		(\bibinfo {year} {2014})}\BibitemShut {NoStop}%
	\bibitem [{\citenamefont {Houari}\ \emph {et~al.}(2018)\citenamefont {Houari},
		\citenamefont {Matar},\ and\ \citenamefont {Eyert}}]{Ni-N_Houari_Matar18}%
	\BibitemOpen
	\bibfield  {author} {\bibinfo {author} {\bibfnamefont {A.}~\bibnamefont
			{Houari}}, \bibinfo {author} {\bibfnamefont {S.~F.}\ \bibnamefont {Matar}}, \
		and\ \bibinfo {author} {\bibfnamefont {V.}~\bibnamefont {Eyert}},\
	}\href@noop {} {\bibfield  {journal} {\bibinfo  {journal} {Electronic
				Structure}\ }\textbf {\bibinfo {volume} {1}},\ \bibinfo {pages} {015002}
		(\bibinfo {year} {2018})}\BibitemShut {NoStop}%
	\bibitem [{\citenamefont {Ma}\ \emph {et~al.}(2018)\citenamefont {Ma},
		\citenamefont {Li}, \citenamefont {Li}, \citenamefont {Li},\ and\
		\citenamefont {Zhang}}]{Mat_Lat18_Ni2N_Zhi_yuan}%
	\BibitemOpen
	\bibfield  {author} {\bibinfo {author} {\bibfnamefont {Z.}~\bibnamefont
			{Ma}}, \bibinfo {author} {\bibfnamefont {Z.}~\bibnamefont {Li}}, \bibinfo
		{author} {\bibfnamefont {S.}~\bibnamefont {Li}}, \bibinfo {author}
		{\bibfnamefont {P.}~\bibnamefont {Li}}, \ and\ \bibinfo {author}
		{\bibfnamefont {H.}~\bibnamefont {Zhang}},\ }\href@noop {} {\bibfield
		{journal} {\bibinfo  {journal} {Materials Letters}\ }\textbf {\bibinfo
			{volume} {229}},\ \bibinfo {pages} {148} (\bibinfo {year}
		{2018})}\BibitemShut {NoStop}%
	\bibitem [{\citenamefont {Niwa}\ \emph {et~al.}(2019)\citenamefont {Niwa},
		\citenamefont {Fukui}, \citenamefont {Terabe}, \citenamefont {Kawada},
		\citenamefont {Kato}, \citenamefont {Sasaki}, \citenamefont {Soda},\ and\
		\citenamefont {Hasegawa}}]{NiN2_InorgChem19}%
	\BibitemOpen
	\bibfield  {author} {\bibinfo {author} {\bibfnamefont {K.}~\bibnamefont
			{Niwa}}, \bibinfo {author} {\bibfnamefont {R.}~\bibnamefont {Fukui}},
		\bibinfo {author} {\bibfnamefont {T.}~\bibnamefont {Terabe}}, \bibinfo
		{author} {\bibfnamefont {T.}~\bibnamefont {Kawada}}, \bibinfo {author}
		{\bibfnamefont {D.}~\bibnamefont {Kato}}, \bibinfo {author} {\bibfnamefont
			{T.}~\bibnamefont {Sasaki}}, \bibinfo {author} {\bibfnamefont
			{K.}~\bibnamefont {Soda}}, \ and\ \bibinfo {author} {\bibfnamefont
			{M.}~\bibnamefont {Hasegawa}},\ }\href@noop {} {\bibfield  {journal}
		{\bibinfo  {journal} {European Journal of Inorganic Chemistry}\ } (\bibinfo
		{year} {2019})}\BibitemShut {NoStop}%
	\bibitem [{\citenamefont {Gajbhiye}\ \emph {et~al.}(2002)\citenamefont
		{Gajbhiye}, \citenamefont {Ningthoujam},\ and\ \citenamefont
		{Weissm{\"u}ller}}]{Gajbhiye2002_Ni-N}%
	\BibitemOpen
	\bibfield  {author} {\bibinfo {author} {\bibfnamefont {N.}~\bibnamefont
			{Gajbhiye}}, \bibinfo {author} {\bibfnamefont {R.}~\bibnamefont
			{Ningthoujam}}, \ and\ \bibinfo {author} {\bibfnamefont {J.}~\bibnamefont
			{Weissm{\"u}ller}},\ }\href@noop {} {\bibfield  {journal} {\bibinfo
			{journal} {physica status solidi (a)}\ }\textbf {\bibinfo {volume} {189}},\
		\bibinfo {pages} {691} (\bibinfo {year} {2002})}\BibitemShut {NoStop}%
	\bibitem [{\citenamefont {Popovi{\'c}}\ \emph {et~al.}(2009)\citenamefont
		{Popovi{\'c}}, \citenamefont {Bogdanov}, \citenamefont {Gonci{\'c}},
		\citenamefont {{\v{S}}trbac},\ and\ \citenamefont
		{Rako{\v{c}}evi{\'c}}}]{ASS09_Popovic_Ni3N}%
	\BibitemOpen
	\bibfield  {author} {\bibinfo {author} {\bibfnamefont {N.}~\bibnamefont
			{Popovi{\'c}}}, \bibinfo {author} {\bibfnamefont {{\v{Z}}.}~\bibnamefont
			{Bogdanov}}, \bibinfo {author} {\bibfnamefont {B.}~\bibnamefont
			{Gonci{\'c}}}, \bibinfo {author} {\bibfnamefont {S.}~\bibnamefont
			{{\v{S}}trbac}}, \ and\ \bibinfo {author} {\bibfnamefont {Z.}~\bibnamefont
			{Rako{\v{c}}evi{\'c}}},\ }\href@noop {} {\bibfield  {journal} {\bibinfo
			{journal} {Applied Surface Science}\ }\textbf {\bibinfo {volume} {255}},\
		\bibinfo {pages} {4027} (\bibinfo {year} {2009})}\BibitemShut {NoStop}%
	\bibitem [{\citenamefont {Phase}\ \emph {et~al.}(2014)\citenamefont {Phase},
		\citenamefont {Gupta}, \citenamefont {Potdar}, \citenamefont {Behera},
		\citenamefont {Sah},\ and\ \citenamefont {Gupta}}]{XAS_beamline}%
	\BibitemOpen
	\bibfield  {author} {\bibinfo {author} {\bibfnamefont {D.~M.}\ \bibnamefont
			{Phase}}, \bibinfo {author} {\bibfnamefont {M.}~\bibnamefont {Gupta}},
		\bibinfo {author} {\bibfnamefont {S.}~\bibnamefont {Potdar}}, \bibinfo
		{author} {\bibfnamefont {L.}~\bibnamefont {Behera}}, \bibinfo {author}
		{\bibfnamefont {R.}~\bibnamefont {Sah}}, \ and\ \bibinfo {author}
		{\bibfnamefont {A.}~\bibnamefont {Gupta}},\ }\href {\doibase
		http://dx.doi.org/10.1063/1.4872719} {\bibfield  {journal} {\bibinfo
			{journal} {AIP Conference Proceedings}\ }\textbf {\bibinfo {volume} {1591}},\
		\bibinfo {pages} {685} (\bibinfo {year} {2014})}\BibitemShut {NoStop}%
	\bibitem [{\citenamefont {Stahn}\ and\ \citenamefont
		{Glavic}(2016)}]{Amor16_cite_1}%
	\BibitemOpen
	\bibfield  {author} {\bibinfo {author} {\bibfnamefont {J.}~\bibnamefont
			{Stahn}}\ and\ \bibinfo {author} {\bibfnamefont {A.}~\bibnamefont {Glavic}},\
	}\href@noop {} {\bibfield  {journal} {\bibinfo  {journal} {Nuclear
				Instruments and Methods in Physics Research Section A: Accelerators,
				Spectrometers, Detectors and Associated Equipment}\ }\textbf {\bibinfo
			{volume} {821}},\ \bibinfo {pages} {44} (\bibinfo {year} {2016})}\BibitemShut
	{NoStop}%
	\bibitem [{\citenamefont {Stahn}\ and\ \citenamefont
		{Glavic}(2017)}]{Amor17_cite_2}%
	\BibitemOpen
	\bibfield  {author} {\bibinfo {author} {\bibfnamefont {J.}~\bibnamefont
			{Stahn}}\ and\ \bibinfo {author} {\bibfnamefont {A.}~\bibnamefont {Glavic}},\
	}in\ \href@noop {} {\emph {\bibinfo {booktitle} {Journal of Physics:
				Conference Series}}},\ Vol.\ \bibinfo {volume} {862}\ (\bibinfo
	{organization} {IOP Publishing},\ \bibinfo {year} {2017})\ p.\ \bibinfo
	{pages} {012007}\BibitemShut {NoStop}%
	\bibitem [{\citenamefont {Braun}(7 99)}]{Parratt32}%
	\BibitemOpen
	\bibfield  {author} {\bibinfo {author} {\bibfnamefont {C.}~\bibnamefont
			{Braun}},\ }\href@noop {} {\emph {\bibinfo {title} {Parratt32- The
				Reflectivity Tool}}}\ (\bibinfo  {publisher} {HMI Berlin},\ \bibinfo {year}
	{1997-99})\BibitemShut {NoStop}%
	\bibitem [{\citenamefont {Gupta}\ \emph {et~al.}(2015)\citenamefont {Gupta},
		\citenamefont {Pandey}, \citenamefont {Tayal},\ and\ \citenamefont
		{Gupta}}]{CoN_AIP_Adv2015}%
	\BibitemOpen
	\bibfield  {author} {\bibinfo {author} {\bibfnamefont {R.}~\bibnamefont
			{Gupta}}, \bibinfo {author} {\bibfnamefont {N.}~\bibnamefont {Pandey}},
		\bibinfo {author} {\bibfnamefont {A.}~\bibnamefont {Tayal}}, \ and\ \bibinfo
		{author} {\bibfnamefont {M.}~\bibnamefont {Gupta}},\ }\href {\doibase
		http://dx.doi.org/10.1063/1.4930977} {\bibfield  {journal} {\bibinfo
			{journal} {AIP Advances}\ }\textbf {\bibinfo {volume} {5}},\ \bibinfo {eid}
		{097131} (\bibinfo {year} {2015})}\BibitemShut {NoStop}%
	\bibitem [{\citenamefont {Kawamura}\ \emph {et~al.}(2000)\citenamefont
		{Kawamura}, \citenamefont {Abe},\ and\ \citenamefont
		{Sasaki}}]{Ni-N_fil_Ts_Kawamura_Vac2000}%
	\BibitemOpen
	\bibfield  {author} {\bibinfo {author} {\bibfnamefont {M.}~\bibnamefont
			{Kawamura}}, \bibinfo {author} {\bibfnamefont {Y.}~\bibnamefont {Abe}}, \
		and\ \bibinfo {author} {\bibfnamefont {K.}~\bibnamefont {Sasaki}},\
	}\href@noop {} {\bibfield  {journal} {\bibinfo  {journal} {Vacuum}\ }\textbf
		{\bibinfo {volume} {59}},\ \bibinfo {pages} {721} (\bibinfo {year}
		{2000})}\BibitemShut {NoStop}%
	\bibitem [{\citenamefont {Ma}\ \emph {et~al.}(2017)\citenamefont {Ma},
		\citenamefont {Zhang}, \citenamefont {Sun}, \citenamefont {Guo},\ and\
		\citenamefont {Li}}]{ASS17_Ni2N_Zhi_yuan}%
	\BibitemOpen
	\bibfield  {author} {\bibinfo {author} {\bibfnamefont {Z.-y.}\ \bibnamefont
			{Ma}}, \bibinfo {author} {\bibfnamefont {H.}~\bibnamefont {Zhang}}, \bibinfo
		{author} {\bibfnamefont {X.}~\bibnamefont {Sun}}, \bibinfo {author}
		{\bibfnamefont {J.}~\bibnamefont {Guo}}, \ and\ \bibinfo {author}
		{\bibfnamefont {Z.-c.}\ \bibnamefont {Li}},\ }\href@noop {} {\bibfield
		{journal} {\bibinfo  {journal} {Applied Surface Science}\ }\textbf {\bibinfo
			{volume} {420}},\ \bibinfo {pages} {196} (\bibinfo {year}
		{2017})}\BibitemShut {NoStop}%
	\bibitem [{\citenamefont {Ito}\ \emph {et~al.}(2015)\citenamefont {Ito},
		\citenamefont {Toko}, \citenamefont {Takeda}, \citenamefont {Saitoh},
		\citenamefont {Oguchi}, \citenamefont {Suemasu},\ and\ \citenamefont
		{Kimura}}]{JAP_15_K_Ito_XAS}%
	\BibitemOpen
	\bibfield  {author} {\bibinfo {author} {\bibfnamefont {K.}~\bibnamefont
			{Ito}}, \bibinfo {author} {\bibfnamefont {K.}~\bibnamefont {Toko}}, \bibinfo
		{author} {\bibfnamefont {Y.}~\bibnamefont {Takeda}}, \bibinfo {author}
		{\bibfnamefont {Y.}~\bibnamefont {Saitoh}}, \bibinfo {author} {\bibfnamefont
			{T.}~\bibnamefont {Oguchi}}, \bibinfo {author} {\bibfnamefont
			{T.}~\bibnamefont {Suemasu}}, \ and\ \bibinfo {author} {\bibfnamefont
			{A.}~\bibnamefont {Kimura}},\ }\href {\doibase
		http://dx.doi.org/10.1063/1.4921431} {\bibfield  {journal} {\bibinfo
			{journal} {Journal of Applied Physics}\ }\textbf {\bibinfo {volume} {117}},\
		\bibinfo {eid} {193906} (\bibinfo {year} {2015}),\
		http://dx.doi.org/10.1063/1.4921431}\BibitemShut {NoStop}%
	\bibitem [{\citenamefont {Kauffman}\ \emph {et~al.}(2016)\citenamefont
		{Kauffman}, \citenamefont {Alfonso}, \citenamefont {Tafen}, \citenamefont
		{Lekse}, \citenamefont {Wang}, \citenamefont {Deng}, \citenamefont {Lee},
		\citenamefont {Jang}, \citenamefont {Lee}, \citenamefont {Kumar} \emph
		{et~al.}}]{Ni_XAS_+2state}%
	\BibitemOpen
	\bibfield  {author} {\bibinfo {author} {\bibfnamefont {D.~R.}\ \bibnamefont
			{Kauffman}}, \bibinfo {author} {\bibfnamefont {D.}~\bibnamefont {Alfonso}},
		\bibinfo {author} {\bibfnamefont {D.~N.}\ \bibnamefont {Tafen}}, \bibinfo
		{author} {\bibfnamefont {J.}~\bibnamefont {Lekse}}, \bibinfo {author}
		{\bibfnamefont {C.}~\bibnamefont {Wang}}, \bibinfo {author} {\bibfnamefont
			{X.}~\bibnamefont {Deng}}, \bibinfo {author} {\bibfnamefont {J.}~\bibnamefont
			{Lee}}, \bibinfo {author} {\bibfnamefont {H.}~\bibnamefont {Jang}}, \bibinfo
		{author} {\bibfnamefont {J.-s.}\ \bibnamefont {Lee}}, \bibinfo {author}
		{\bibfnamefont {S.}~\bibnamefont {Kumar}},  \emph {et~al.},\ }\href@noop {}
	{\bibfield  {journal} {\bibinfo  {journal} {ACS Catalysis}\ }\textbf
		{\bibinfo {volume} {6}},\ \bibinfo {pages} {1225} (\bibinfo {year}
		{2016})}\BibitemShut {NoStop}%
	\bibitem [{\citenamefont {Blundell}\ and\ \citenamefont
		{Bland}()}]{PNR92_Blundell}%
	\BibitemOpen
	\bibfield  {author} {\bibinfo {author} {\bibfnamefont {S.}~\bibnamefont
			{Blundell}}\ and\ \bibinfo {author} {\bibfnamefont {J.}~\bibnamefont
			{Bland}},\ }\href@noop {} {\bibfield  {journal} {\bibinfo  {journal}
			{Physical Review B}\ }\textbf {\bibinfo {volume} {46}}}\BibitemShut {NoStop}%
	\bibitem [{\citenamefont {Ott}(2011)}]{SimulReflec}%
	\BibitemOpen
	\bibfield  {author} {\bibinfo {author} {\bibfnamefont {F.}~\bibnamefont
			{Ott}},\ }\href
	{\\http://www-llb.cea.fr/prism/programs/simulreflec/simulreflec.html}
	{\bibfield  {journal} {\bibinfo  {journal} {SIMULREFLEC}\ } (\bibinfo {year}
		{V1.7 2011})}\BibitemShut {NoStop}%
	\bibitem [{\citenamefont {Linnik}\ \emph {et~al.}(2013)\citenamefont {Linnik},
		\citenamefont {Prudnikov}, \citenamefont {Shalaev}, \citenamefont {Linnik},
		\citenamefont {Varyukhin}, \citenamefont {Kostyrya},\ and\ \citenamefont
		{Burkhovetskii}}]{Mag_NiN_Linnik2013}%
	\BibitemOpen
	\bibfield  {author} {\bibinfo {author} {\bibfnamefont {A.}~\bibnamefont
			{Linnik}}, \bibinfo {author} {\bibfnamefont {A.}~\bibnamefont {Prudnikov}},
		\bibinfo {author} {\bibfnamefont {R.}~\bibnamefont {Shalaev}}, \bibinfo
		{author} {\bibfnamefont {T.}~\bibnamefont {Linnik}}, \bibinfo {author}
		{\bibfnamefont {V.}~\bibnamefont {Varyukhin}}, \bibinfo {author}
		{\bibfnamefont {S.}~\bibnamefont {Kostyrya}}, \ and\ \bibinfo {author}
		{\bibfnamefont {V.}~\bibnamefont {Burkhovetskii}},\ }\href@noop {} {\bibfield
		{journal} {\bibinfo  {journal} {Technical Physics Letters}\ }\textbf
		{\bibinfo {volume} {39}},\ \bibinfo {pages} {143} (\bibinfo {year}
		{2013})}\BibitemShut {NoStop}%
	\bibitem [{\citenamefont {Shalayev}\ \emph {et~al.}(2014)\citenamefont
		{Shalayev}, \citenamefont {Prudnikov}, \citenamefont {Kutrovskaya},
		\citenamefont {Varyukhin}, \citenamefont {Linnik},\ and\ \citenamefont
		{Arakelian}}]{Ni4N_high_Ts_Shalayev2014}%
	\BibitemOpen
	\bibfield  {author} {\bibinfo {author} {\bibfnamefont {R.}~\bibnamefont
			{Shalayev}}, \bibinfo {author} {\bibfnamefont {A.}~\bibnamefont {Prudnikov}},
		\bibinfo {author} {\bibfnamefont {S.}~\bibnamefont {Kutrovskaya}}, \bibinfo
		{author} {\bibfnamefont {V.}~\bibnamefont {Varyukhin}}, \bibinfo {author}
		{\bibfnamefont {A.}~\bibnamefont {Linnik}}, \ and\ \bibinfo {author}
		{\bibfnamefont {S.}~\bibnamefont {Arakelian}},\ }\href@noop {} {\bibfield
		{journal} {\bibinfo  {journal} {Functional Materials}\ } (\bibinfo {year}
		{2014})}\BibitemShut {NoStop}%
\end{thebibliography}
%

\end{document}